\documentclass[a4paper,fleqn,usenatbib]{mnras}
\usepackage{graphics,graphicx,rotating}
\usepackage{natbib}
\usepackage{xspace}
\usepackage{amssymb}
\usepackage{amsmath}
\usepackage{epstopdf}
\usepackage{subfigure}
\usepackage{pdflscape}
\usepackage{rotating}
\usepackage{xcolor}
\usepackage{xspace}
\usepackage{amssymb}
\usepackage{amsmath}
\usepackage{epstopdf}
\usepackage{subfigure}
\usepackage{xcolor}
\usepackage[percent]{overpic}
\usepackage{xcolor}
\usepackage{pict2e}
\usepackage[T1]{fontenc}
\usepackage{lmodern}
\newcommand{\mdef}{{$M_{\rm def}$}}
\newcommand{\sm}{$\sim~$}
\newcommand{\fc}{$\farcs$}
\newcommand{\sn}{$M_{\sun}$}
\newcommand{\mstar}{$M_{*}$}
\newcommand{\dg}{$^{\circ}$}
\begin{document}

\title [A high-resolution, multi-band analysis of NGC 5322.]{The
  nuclear  activity and central structure of the elliptical galaxy NGC
  5322}
\author[Dullo et al.]{
         Bililign T.\  Dullo,$^{1,2,3}$\thanks{bdullo@ucm.es},
         Johan  H.  Knapen$^{2,3}$, 
         David R. A. Williams$^{4}$,
        Robert J. 
     \newauthor   Beswick$^{5}$,
         George Bendo$^{5}$,
         Ranieri D. Baldi$^{4}$,
         Megan Argo$^{5,6}$,
         Ian M. McHardy$^{4}$,
         \newauthor  Tom Muxlow$^{5}$,
         J. Westcott$^{7}$\\
$^{1}$Departamento de   Astrof\'isica y Ciencias de
    la Atm\'osfera, Universidad Complutense de Madrid, E-28040 Madrid,
    Spain\\
$^{2}$Instituto de Astrof\'isica de Canarias, V\'ia
  L\'actea S/N, E-38205, La Laguna, Tenerife, Spain\\
$^{3}$Departamento de Astrof\'isica, Universidad de
  La Laguna, E-38206, La Laguna, Tenerife, Spain\\
$^{4}$School of Physics and Astronomy, University of Southampton, Southampton, SO17 1BJ, UK\\
$^{5}$Jodrell Bank Centre for Astrophysics, School of
  Physics and Astronomy, The University of Manchester, Alan Turing
  Building, Oxford Road,\\
 Manchester, M13 9PL, UK\\
$^{6}$Jeremiah Horrocks Institute, University of Central Lancashire, Preston PR1 2HE, UK\\
$^{7}$Centre for Astrophysics Research, University of Hertfordshire, College Lane, Hatfield, AL10 9AB, UK}
\maketitle
\label{firstpage}
\begin{abstract}

  We have analysed a new high-resolution e-MERLIN 1.5 GHz radio
  continuum map together with {\it HST} and SDSS imaging of NGC 5322,
  an elliptical galaxy hosting radio jets, aiming to understand the
  galaxy's central structure and its connection to the nuclear
  activity. We decomposed the composite {\it HST} + SDSS surface
  brightness profile of the galaxy into an inner stellar disc, a
  spheroid, and an outer stellar halo. Past works showed that this
  embedded disc counter-rotates rapidly with respect to the
  spheroid. The {\it HST} images reveal an edge-on nuclear dust disc
  across the centre, aligned along the major-axis of the galaxy and
  nearly perpendicular to the radio jets. After careful masking of
  this dust disc, we find a central stellar mass deficit $M_{\rm def}$
  in the spheroid, scoured by SMBH binaries with final mass
  $M_{\rm BH}$ such that $M_{\rm def}/M_{\rm BH} \sim 1.3 - 3.4$.  We
  propose a three-phase formation scenario for NGC~5322 where a few
  ($2-7$) `dry' major mergers involving SMBHs built the spheroid with
  a depleted core. The cannibalism of a gas-rich satellite
  subsequently creates the faint counter-rotating disc and funnels
  gaseous material directly onto the AGN, powering the radio core with
  a brightness temperature of
  $T_{\rm B,core} \sim 4.5 \times 10^{7}$~K and the low-power radio
  jets ($P_{\rm jets}\sim 7.04 \times 10^{20}$~W~Hz$^{-1}$) which
  extend $\sim 1.6$ kpc. The outer halo can later grow via minor
  mergers and the accretion of tidal debris. The low-luminosity
  AGN/jet-driven feedback may have quenched the late-time nuclear star
  formation promptly, which could otherwise have replenished the
  depleted core.

\end{abstract}


\section{Introduction}

Elliptical galaxies are thought to have formed hierarchically via
mergers of smaller systems and accretion events (e.g.,
\citealt{1978MNRAS.183..341W}; \citealt{1994MNRAS.271..781C};
\citealt{2001ApJ...561..517K}; \citealt{2005Natur.435..629S};
\citealt{2006ApJ...636L..81N}; \citealt{2009ApJS..181..486H};
\citealt{2017MNRAS.470.3507M}).  Merger-driven inflow of gas was
invoked to produce nuclear starbursts or fuel the central black
hole. The latter may drive powerful relativistic jets that can heat the
surrounding gas, and as such impact on the central stellar structures
in elliptical galaxies (e.g., \citealt{1986ApJ...310..593C};
\citealt{1992ApJ...400..460H, 1993ApJ...409..548H};
\citealt{1994ApJ...431L...9M}; \citealt{1995ApJ...452L..91B};
\citealt{2001AJ....122.2791S}; \citealt{2006MNRAS.365...11C};
\citealt{2014MNRAS.439.2291W};
\citealt{2015ARA&A..53...51S}). Observations reveal that elliptical
galaxies possess distinct structures on scales of a few kiloparsecs
down to tens of parsecs such as depleted cores, nuclear star clusters,
small-scale stellar discs, nuclear dust discs/lanes, and AGN (e.g.,
\citealt{1993AJ....106.1371C}; \citealt{1994ESOC...49..147K};
\citealt{1994AJ....108.1567J}; \citealt{1994AJ....108.1598F};
\citealt{1995AJ....110.2622L}; \citealt{1997ApJ...481..710C},
\citealt{1997AJ....114.1771F}; \citealt{2000ApJ...529...93K};
\citealt{2001AJ....121.2431R}; \citealt{2004AJ....127.1917T};
\citealt{2005AJ....129.2138L, 2007ApJ...662..808L};
\citealt{2006ApJS..165...57C}; \citealt{2006ApJS..164..334F};
\citealt{2009ApJS..182..216K}; \citealt{2012ApJ...755..163D,
  2014MNRAS.444.2700D}; \citealt{2012ApJS..203....5T};
\citealt{2016ApJS..222...10S};
\citealt{2017arXiv170702277D}). However, the interplay between such
central structures and the nuclear activity of the galaxies remain
largely unclear.

Luminous elliptical galaxies brighter than
$M_{\rm B} \sim -20.5 \pm 0.05$ mag contain partially-depleted cores,
with sizes typically 20 $-$ 500 pc, that are thought to be scoured by
binary supermassive black holes formed in gas-poor (`dry') major
mergers (e.g., \citealt{1980Natur.287..307B};
\citealt{1991Natur.354..212E}; \citealt{1997AJ....114.1771F};
\citealt{2006ApJ...648..976M}). The coexistence of such cores and
rapidly rotating nuclear stellar and dust discs in galaxies is of
interest since it may lead us to understand how such galaxies can
acquire and transfer gaseous material inward to build the nuclear
stellar/dust discs, fuel the central AGN, but not refill the depleted
core with young stars. The radio elliptical (E3$-$4) galaxy NGC 5322
(RC3, \citealt{1991rc3..book.....D}), with a depleted core, i.e., a
central stellar mass deficit (e.g., \citealt{1997ApJ...481..710C};
\citealt{2011MNRAS.415.2158R}; \citealt{2013KRb}), nuclear stellar
disc and LINER-type nuclear emission \citep{2009A&A...508..603B},
is a promising candidate which allows detailed, multi-wavelength
analysis of the central structure, AGN activities and jet structures.

The Legacy e-MERLIN Multi-band Imaging of Nearby Galaxies (LeMMINGs;
\citealt{2015aska.confE..70B}) survey is a high-resolution radio
continuum survey designed, in part, to study such galaxies by
utilising synergies from a large sample of high-resolution,
multi-wavelength data, including from {\it HST}: optical plus IR, {\it
  Spitzer}: IR, {\it Herschel}: IR, {\it Chandra}: X-ray and {\it
  GALEX}: UV.  The project is composed of two complementary radio
surveys: moderately deep snapshots of a large sample of 280 nearby
($3-100$ Mpc) galaxies over a large luminosity, size and morphological
type range, selected from the Palomar Bright Galaxies Survey
(\citealt{1997ApJS..112..315H}), and a deep survey of 6/280
representative galaxies. These radio surveys coupled with the
ancillary data will probe the central structures, AGN/jet structure
and star formation rates in galaxies (see
\citealt{2017MNRAS.467.2113W} and \citealt{2017Williams}). NGC 5322 is
among the first galaxies observed with e-MERLIN at 1.5 GHz as part of
the moderately deep LeMMINGs survey.

Isophotal analysis of ground-based optical images by
\citet{1988A&A...202L...5B} showed that the isophote shapes in NGC
5322 change from discy isophotes at $R \la 8\arcsec$, due to the
galaxy's small-scale, edge-on stellar disc, to boxy isophotes at
larger radii ($8\arcsec \la R\la 35\arcsec$). Embedded discs are
common in elliptical galaxies (e.g., \citealt[references
therein]{2001AJ....121.2431R}; \citealt{2012ApJ...755..163D};
\citealt{2016ApJ...831..132G}).
The \citet{1988A&A...202L...5B} kinematic data for NGC~5322, out to
$35\arcsec$, also revealed that the inner region
$(R \la 2.6 \arcsec)$, where the embedded stellar disc contributes
light significantly, is counter-rotating with respect to the rest of
the galaxy (see also \citealt{1992A&A...258..250B},
\citealt{1992MNRAS.254..389R} and \citealt[their Fig
C4]{2011MNRAS.414.2923K}). This counter-rotating core (KDC) exhibits
enhanced Mg$_{2}$ relative to the outer parts of the galaxy
\citep{1992A&A...258..250B, 1999ApJ...527..573K}. The Mg$_{2}$ line
index is thought to trace primarily metallicity. The stellar
population analysis within $1R_{\rm e}$ by \citet[and references
therein]{2010MNRAS.408..272S} has revealed that the galaxy is very
metal rich ([Z/H] $\sim 0.45$) with a super-solar $\alpha$-abundance
ratio ([$\alpha$/Fe] $\sim$ 0.30) in the central
region. \citet{1997ApJ...481..710C} found that the $V- I$ colour of
the KDC differs only slightly ($\la 0.02$ mag) from that of the
surrounding regions. These isophotal, kinematical and stellar
population properties and the colour of NGC~5322 were attributed to
the build-up of the galaxy through a major merger of two gas-rich
galaxies (e.g., \citealt {1990ApJ...364L..33S};
\citealt{1991Natur.354..210H}; \citealt{1992A&A...258..250B};
\citealt{1992AJ....104.1039S}). However, the presence of a depleted
core in NGC~5322 argues against such a gas-rich, major merger scenario
in which the merger-driven gas dissipation and ensuing nuclear
starburst are thought to produce coreless galaxies.

Reducing the \citet{2001ApJ...546..681T} distance moduli by 0.06 mag
\citep{2002MNRAS.330..443B} yields a distance of 30.3 Mpc for
NGC~5322. The galaxy has a central velocity dispersion $\sigma \sim$
229 km s$^{-1}$ (HyperLeda\footnote{http://leda.univ- lyon1.fr},
\citealt{2003A&A...412...45P}). We assume that H$_{0}$ = 70 km
s$^{-1}$ Mpc$^{-1}$ and that 1 arcsec corresponds to 145 pc.

The aim of this work is to investigate the formation mechanisms for
NGC~5322 using a multi-wavelength analysis of the galaxy's central
structure and the connection to the nuclear activity. To achieve this,
we use our new 1.5 GHz e-MERLIN observations plus {\it HST} and SDSS
imaging. These data and the pertaining data reduction steps are
discussed in Section~\ref{Sec2}.  After carefully masking the galaxy's
nuclear dust disc, for the first time, we decompose the $200\arcsec$
composite ({\it HST} + SDSS) light profile into an inner (exponential)
stellar disc, a core-S\'ersic spheroid and an outer exponential
stellar halo (Section~\ref{Sec3}).  Section~\ref{Sec3} also details
our radio, isophotal and light profile fitting analyses, along with a
literature comparison. Sections \ref{Sec4.1} and \ref{Sec4.2} provide
our measurements of the central stellar mass deficit and merger rate
for NGC~5322.
In Section~\ref{Sec4.4}, we discuss the formation of
NGC~5322. Section~\ref{Conc}
summarizes our main conclusions.

\begin{figure*}
\hspace*{-.88319772599cm}  
\vspace*{-.46772599cm}  
\linethickness{36pt} 
\begin{overpic}[angle=0,scale=0.6762]{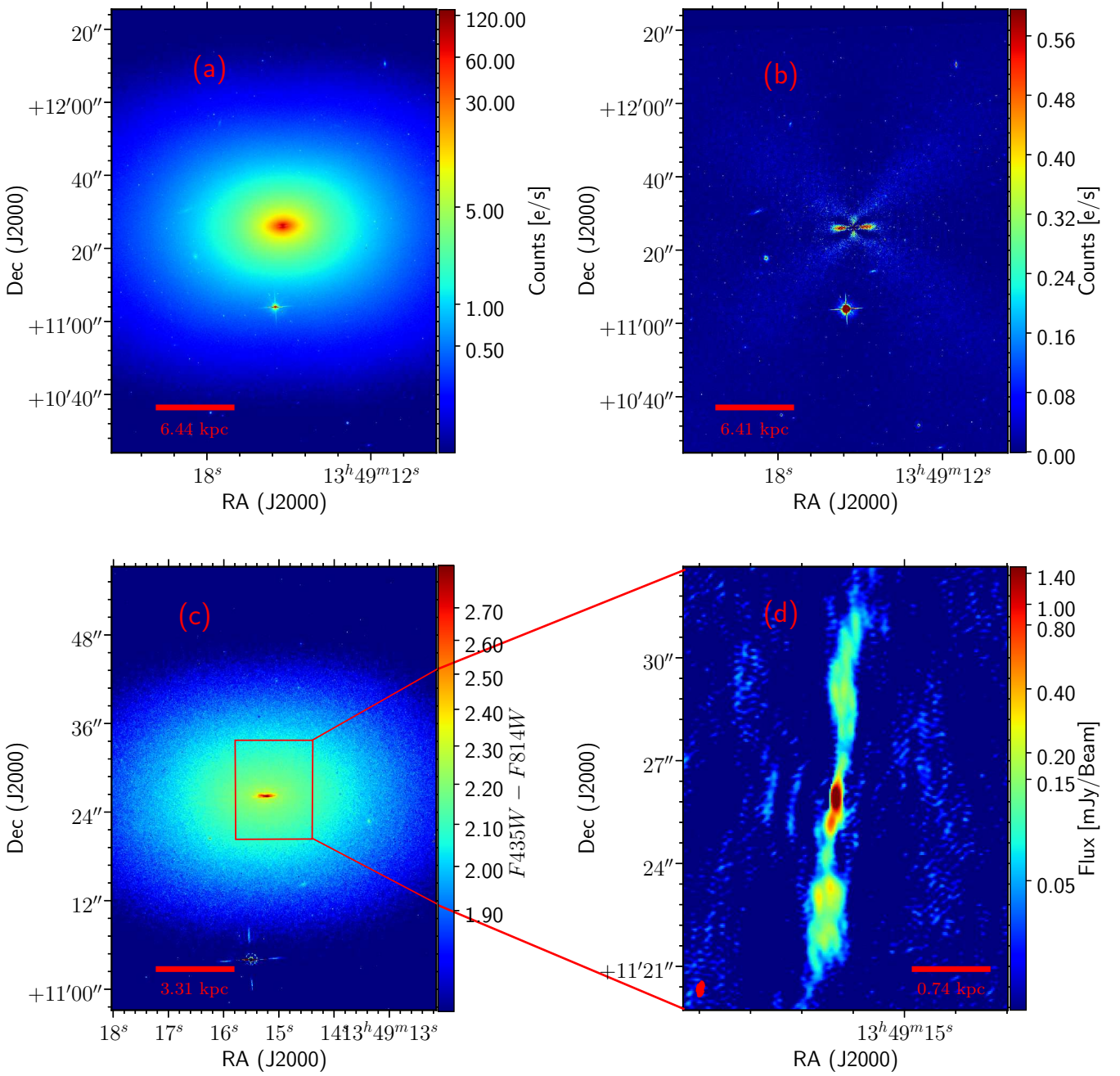} 
\put(16,46.){\color{red}\line(0,1){0.384}}
\put(13.7,44.4){\color{red}{6.44 kpc}}
\put(53,46.){\color{red}\line(0,1){0.384}}
\put(50.75,44.4){\color{red}{6.41 kpc}}
\put(16,8.55){\color{red}\line(0,1){0.384}}
\put(13.7,7.05){\color{red}{3.31 kpc}}
\put(66,8.55){\color{red}\line(0,1){0.384}}
\put(63.75,7.05){\color{red}{0.74 kpc}}
\end{overpic}
\caption{ (a) {\it HST} ACS $F814W$ image of NGC 5322.  (b) ACS
  $F814W$ residual image of NGC 5322 created by subtracting the {\sc
    ellipse} model image from the ACS $F814W$ image.  (c) {\it HST}
  ACS $F435W - F814W$ colour map of the NGC 5322. The relatively
  redder ($F435W - F814W$ $\ga 2.3$) nuclear regions are due to the
nuclear dust disc with radius $R \sim 1\farcs$7 $\approx$ 246.5
  pc and (d) e-MERLIN 1.5 GHz radio continuum map of NGC 5233.  The
  beam is shown in the lower left corner. In all the panels north is
  up, and east is to the left.  The nuclear dust disc is embedded in
  the small-scale stellar disc with a half-light radius
  $R_{\rm e} \sim 356.7$ pc.  Both the dust and stellar discs are
  edge-on and aligned along the major axis of the galaxy, i.e., in the
  east-west direction. The dark regions, bright cones along the
  major-axis and the `X'-shaped structure in the residual image are
  due to the dust disc, stellar disc and boxy spheroid,
  respectively. }
\label{Fig1} 
\end{figure*}

\begin{figure}
 \begin{tabular}{|c|c|}
\hspace*{-.4570619982453901270899cm}   
 \includegraphics[angle=0,scale=0.247963045]{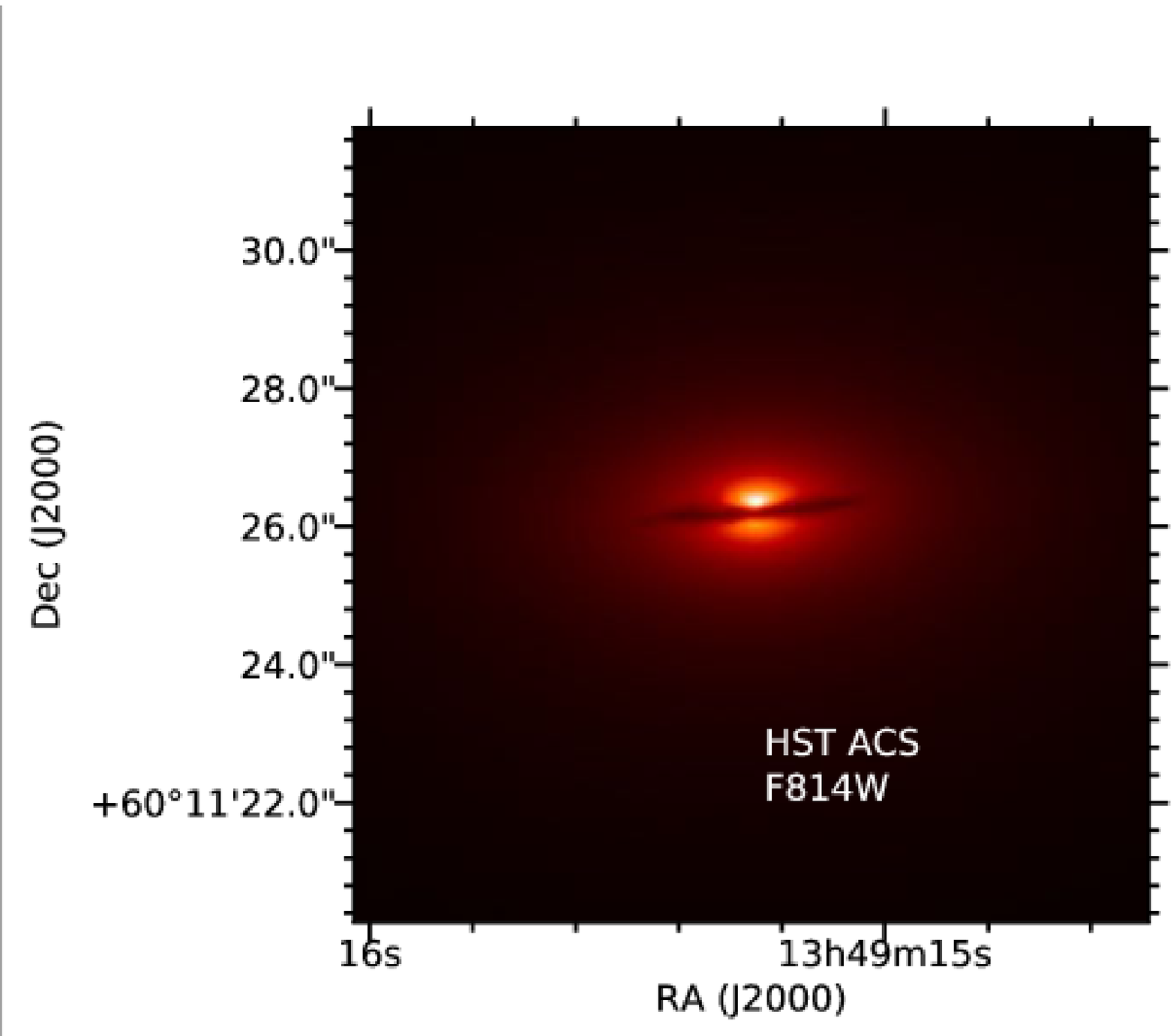}
\hspace*{-1.302847882453901270899cm}   
\includegraphics[angle=0,scale=0.247963045]{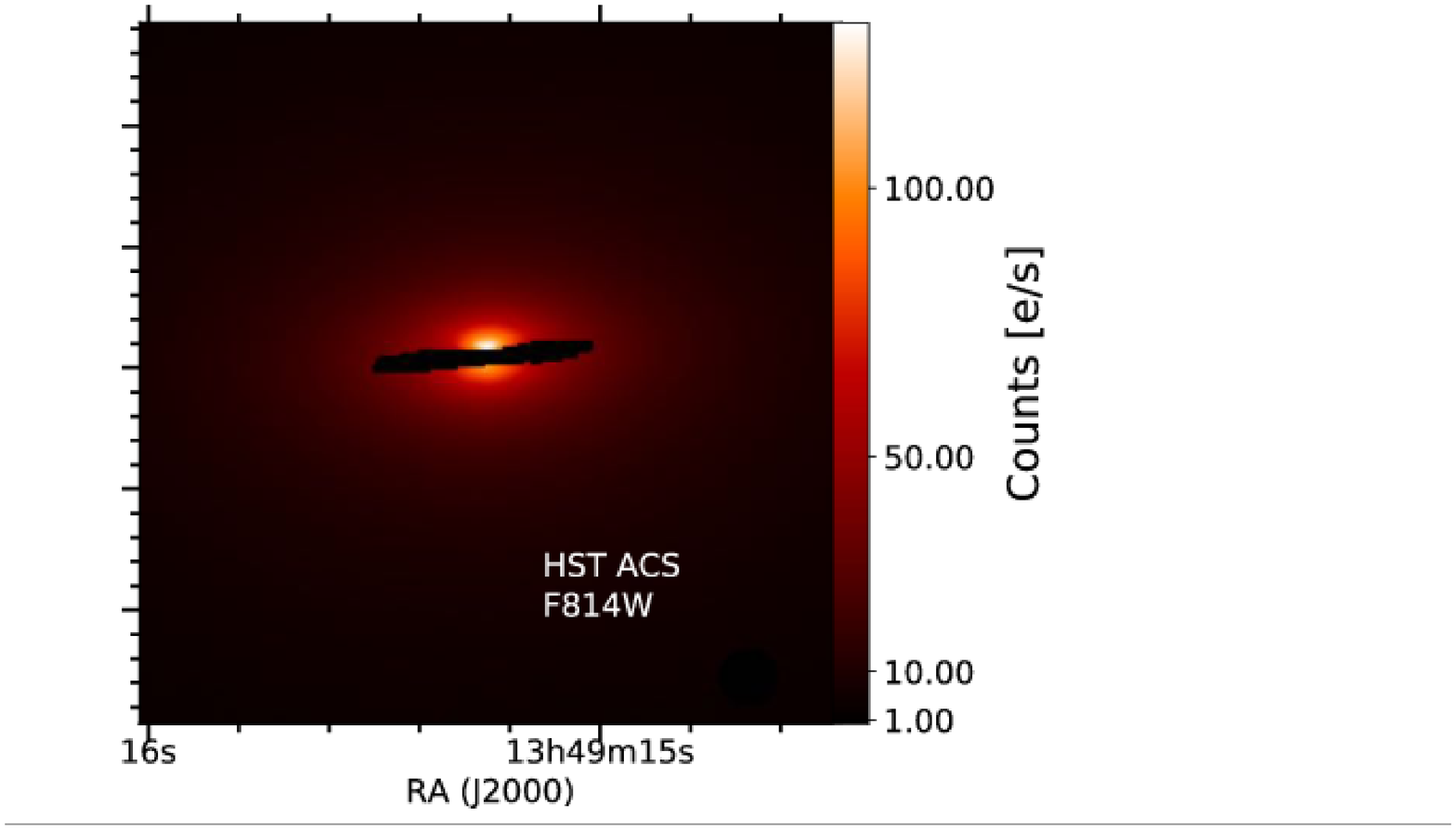}\\
\hspace*{-1.464051301572453901270899cm}   
\vspace*{-.357882453901270899cm}   
 \includegraphics[angle=0,scale=0.2703045]{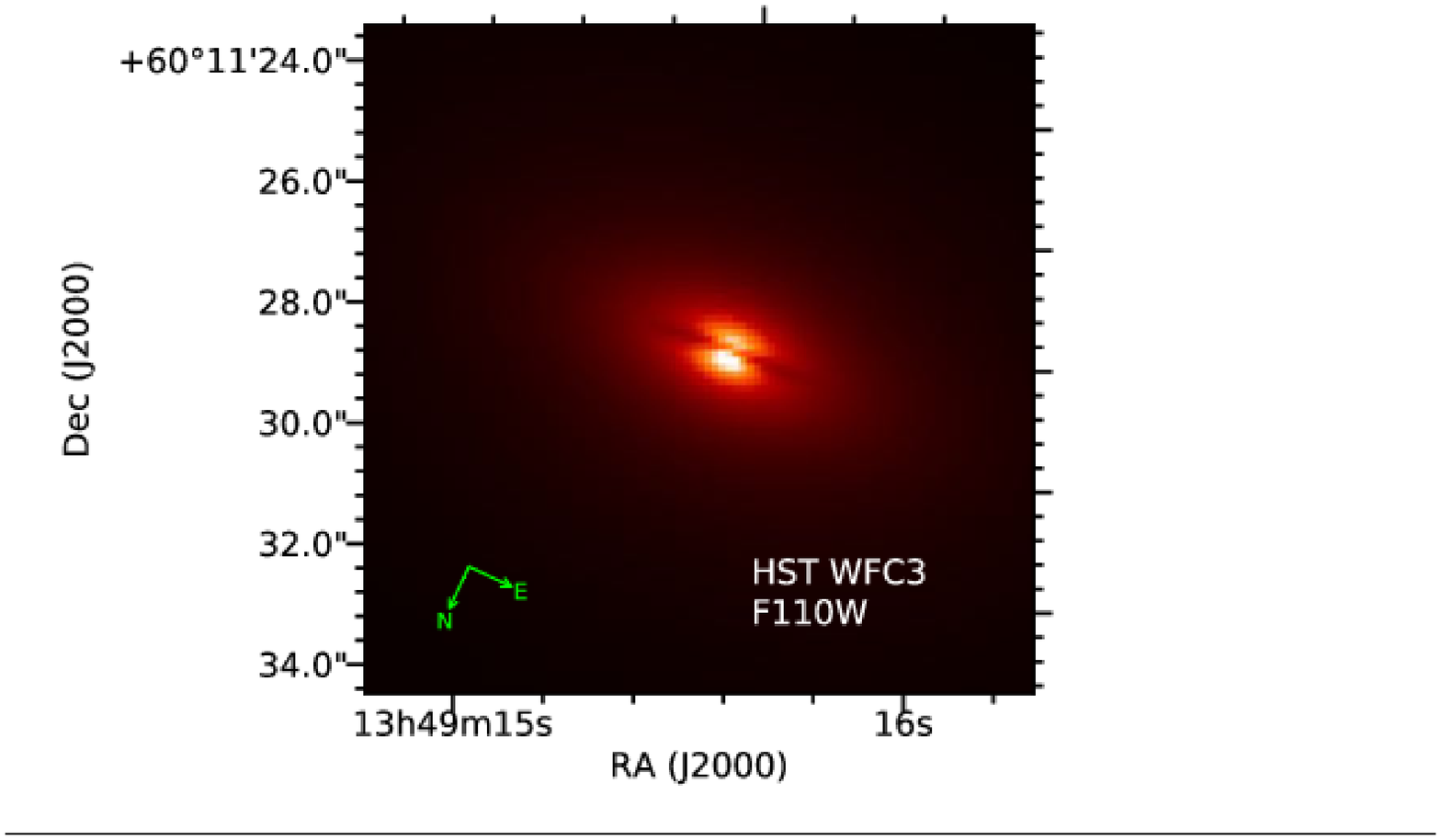}
\hspace*{-1.9037882453901270899cm}   
\includegraphics[angle=0,scale=0.25904045]{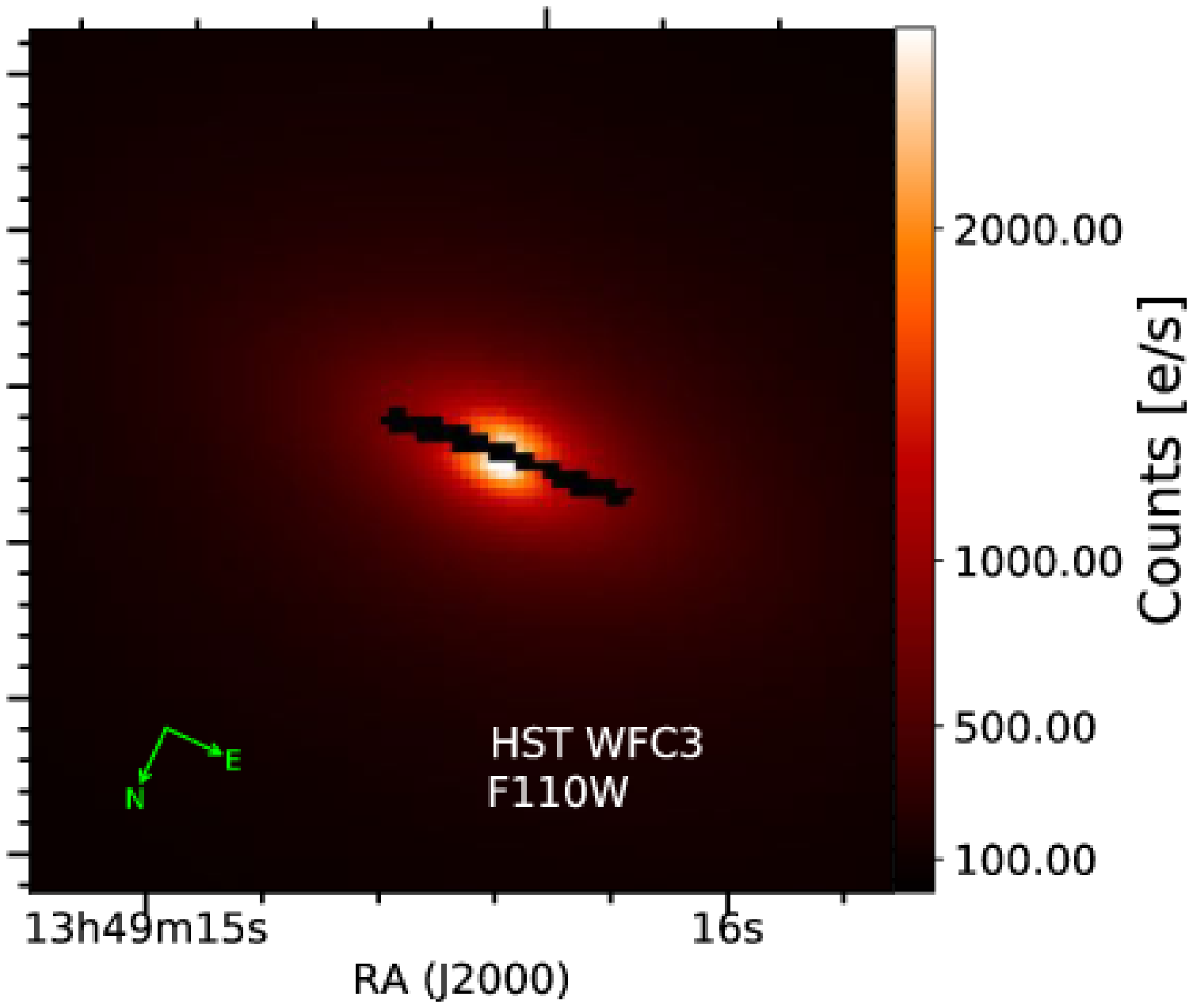}
 \end{tabular}
 \caption{Top panels: {\it HST} ACS F814W image of NGC~5322 showing
   the central regions of the galaxy with the nuclear dust disc (left)
   and the corresponding masked regions (black areas) overplotted on
   the image (right). North is up and east is to the left. Bottom
   panels: as in the top panels but for the {\it HST} WFC3 F110W
   image. The orientation of the WFC3 images is marked with arrows. }
\label{Fig20} 
\end{figure}

\begin{figure*}
\hspace*{-.62590772599cm}   
\vspace*{-.3772599cm}   
\linethickness{36pt}
\begin{overpic}[angle=0,scale=0.569748002]{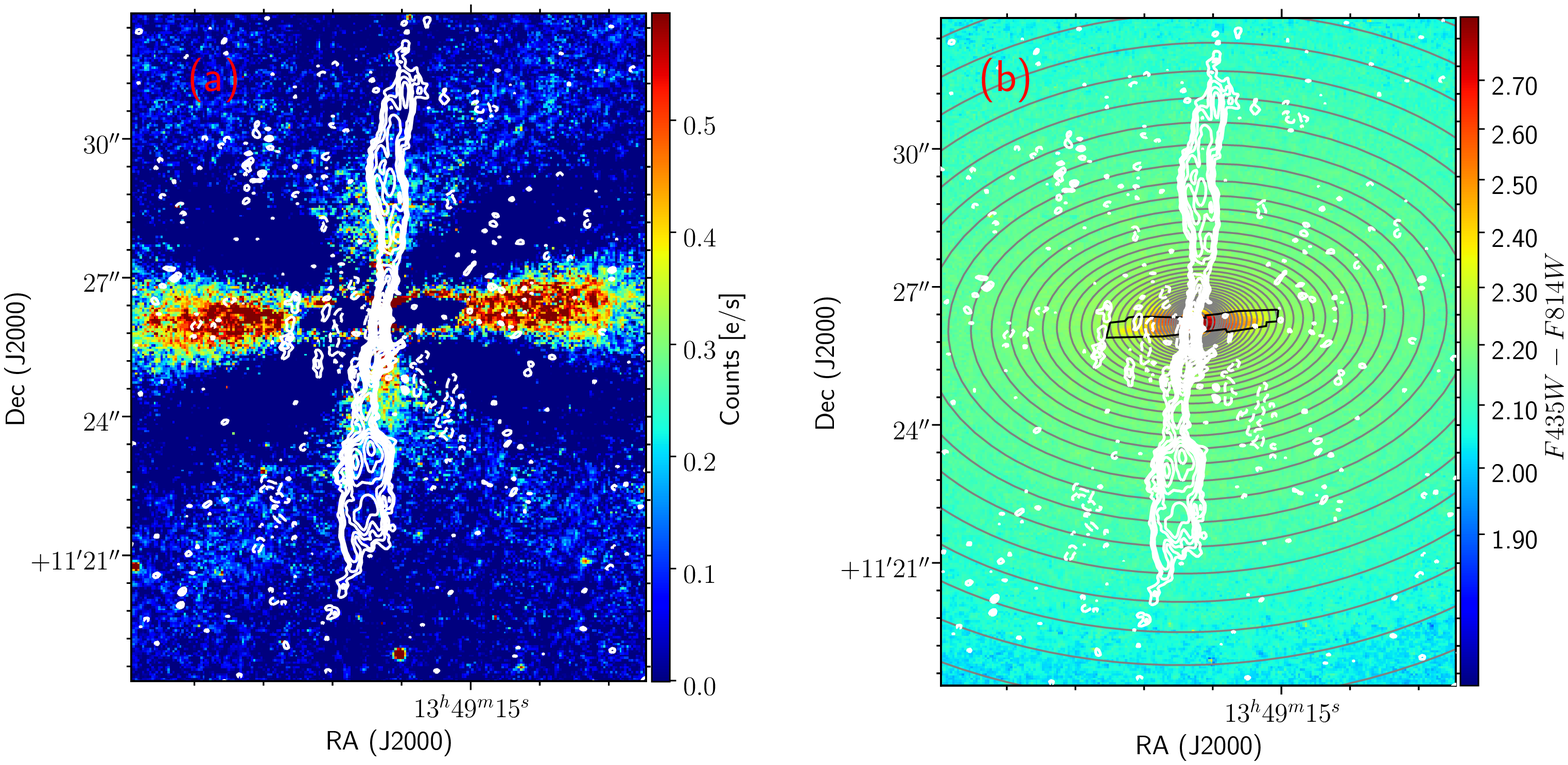} 
\put(16,8.55){\color{red}\line(0,1){0.384}}
\put(13.5,7.05){\color{red}{0.59 kpc}}
\put(64,8.55){\color{red}\line(0,1){0.384}}
\put(61.5,7.05){\color{red}{0.60 kpc}}
\end{overpic}
\caption{e-MERLIN 1.5 GHz contours of NGC 5322 overlaid on the (a)
  residual image and (b) $F435W - F814W$ colour map of the galaxy,
  highlighting the residual structures and colour excess due to the
  dust disc.  We overplot the elliptical isophotes derived from the
  {\sc ellipse} fit to the ACS $F814W$ image (see
  Section~\ref{Sec2.5}) to illustrate the position angle of the galaxy
  relative to the dust disk and radio jets. For clarity we do not show
  a few isophotes inside $R \la 0\farcs4$. While the black contour
  (with a semi-major axis of $\sim 1\farcs7$) encloses the masked
  region shown in Fig.~\ref{Fig20}, we use the ACS $F814W$ light
  profile at $3\arcsec \la R \la 40\arcsec$
  (Section~\ref{Sec2.5}). The radio jet is nearly perpendicular to the
  nuclear stellar and dust discs that are aligned along the major axis
  of the galaxy and result in the bright cones in East-West direction
  (a) and the colour excess (b), respectively. In both panels north is
  up, and east is to the left. The contour levels are (-1.00, 1.00,
  1.41, 2.00, 2.83, 4., 5.66, 8.00, 11.31, 16.00, 22.63, 32.00, 45.25,
  64.00, 90.51, 128.00, 181.00, 256.00, 362.00 and 512.00) $\times$
  3.326E-05 Jy beam$^{-1}$.}
\label{Fig2} 
\end{figure*}

 \section{Observations and data reduction}\label{Sec2}

\subsection{e-MERLIN radio data} 

High-sensitivity, high-resolution radio continuum observations of
NGC\,5322 were carried out using the e-MERLIN array at L-band
(1.25-1.75~GHz) on 17-May-2015 as part of the e-MERLIN LeMMINGs legacy
programme \citep{2014evn..confE..10B}. We used all seven of the
e-MERLIN telescopes throughout these observations, including the 76-m
Lovell telescope. The target source was observed for a total on-source
time of 5.7\,hrs. These observations were regularly interspersed with
short scans of the nearby phase calibration source, J1335+5844, with a
13.2 minute target to phase calibrator cycle. Observations of flux
density calibration source, 3C286, and bandpass calibrator, OQ208,
were made at the beginning and end of the observing run. A total
bandwidth of 512 MHz was correlated into 8 adjacent spectral windows
with each spectral window correlated into 512 frequency channels per
polarisation.  Following initial data editing and the excision of RFI
these data were averaged in frequency (0.5 MHz) and time (2 sec)
before being processed using the e-MERLIN pipeline
(\citealt{2015arXiv150204936A}) to apply standard multi-frequency
calibration procedures to determine and apply the complex gain
solutions derived from the calibration sources. Following these
pipeline procedures a number of self-calibration cycles were carried
out on the target source, NGC\,5322, before reweighting of the
relative sensitivities of the individual e-MERLIN telescopes (as a
function of time and frequency) and final imaging.

Multi-frequency synthesis imaging techniques were applied to the
calibrated target source data using a variety of weighting schemes
optimising for sensitivity and angular resolution. The final images
(see Fig.~\ref{Fig1}d) with a restoring beam of
$0\farcs43\times 0\farcs 17$, a $\sim$ 41 $\times$ 41 arcsec$^{2}$
field-of-view (FOV), rms noise level of 11.6\,$\mu$Jy\,beam$^{-1}$ and
peak target flux density of 5.26\,mJy\,beam$^{-1}$ were used for
subsequent analysis.

\subsection{{\it HST} imaging data} 

{\it HST} Advanced Camera for Surveys (ACS;
\citealt{1998SPIE.3356..234F}) Wide Field Channel (WFC) $F435W$ and
$F814W$ images of NGC~5322 ({\it HST} Proposal ID 9427,
\citealt{2006ApJ...636...90H}) were retrieved from the public Hubble
Legacy Archive (HLA\footnote{https://hla.stsci.edu/hlaview.html}).
The ACS $F435W$ and $F814W$ filters are similar to Johnson's $B$- and
$I$-bands, respectively. In addition, we retrieved the {\it HST} Wide
Field Camera 3 (WFC3) IR $F110W$ image from the Mikulski Archive for
Space Telescopes (MAST\footnote{https://archive.stsci.edu}) to better
account for the edge-on nuclear dust disc visible in the images and in
our colour map (Figs.~\ref{Fig1}, ~\ref{Fig20} and ~\ref{Fig2}). The ACS WFC and
WFC3 IR cameras have pixel scales of 0$\farcs$05 pixel$^{-1}$ and
0$\farcs$13 pixel$^{-1}$, respectively.

\subsection{ SDSS imaging data}\label{sdss}

We obtained the SDSS $i$-band mosaic image of NGC~5322, with a pixel
scale of $0$\farcs$396$ and a $\sim$ 18 $\times$ 18
arcmin$^{-2}$ FOV from the Data Release 13
(DR13\footnote{http://www.sdss.org/dr13/}) database. This image was
used to determine reliably the sky background of NGC~5322.
 
\begin{figure}
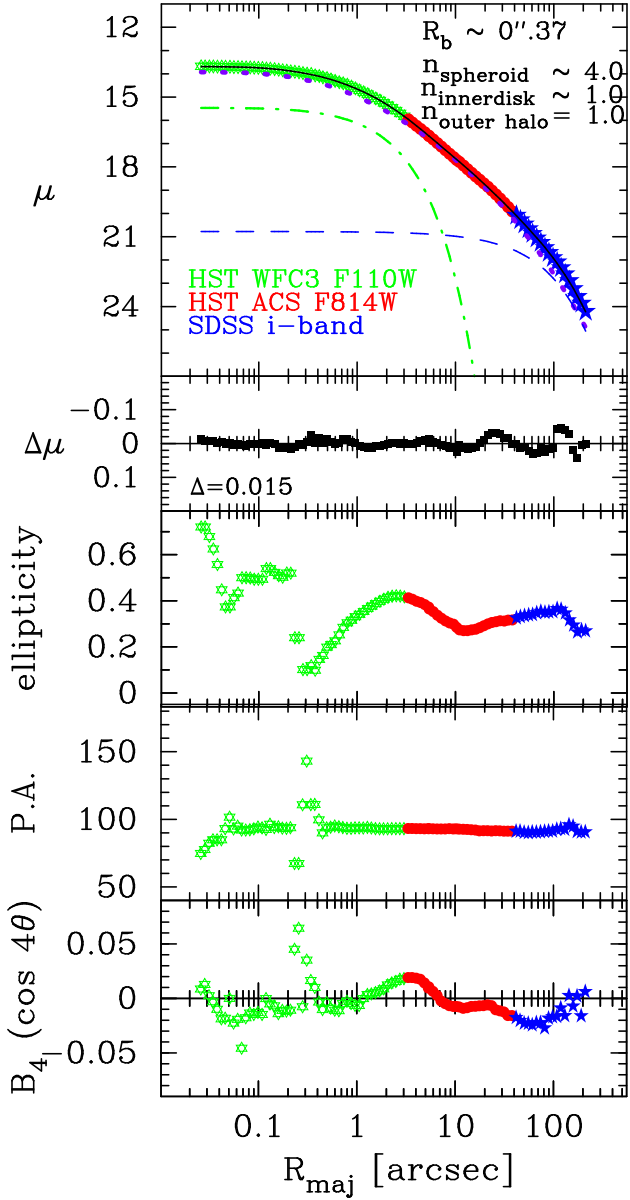

 \hspace*{-.2599cm}   
\vspace*{-.157299cm}   
 \includegraphics[angle=270,scale=0.82]{N5322_Fit.ps}\\
 \hspace*{-.29299cm}   
 \includegraphics[angle=270,scale=0.82]{N5322_prof.ps}
 \caption{Three-component inner disc (dash-dotted green curve) + boxy
   spheroid (dotted magenta curve) + outer halo (dashed blue curve)
   model fit to the composite {\it HST} WFC3 IR $F110W$ + {\it HST}
   ACS $F814W$ + SDSS $i$-band surface brightness profile of NGC 5322
   (see Section~\ref{Sec2.5}). The solid black curve shows the
   complete fit to the profile. The WFC3 $F110W$ and SDSS $i$-band
   light profiles were shifted to match the ACS $F814W$ data. This
   decomposition yields a depleted core with a break radius of
   $R_{\rm b}\sim 0\farcs37 \pm 0\farcs04$. Note that we extracted the
   innermost ($R \la 3\arcsec$) light profile of the galaxy using (i)
   the WFC3 IR $F110W$ image and (iii) a careful mask to avoid the
   nuclear dust disc, with a thickness of $\sim 0\farcs$2
   $\approx$ 28 pc near the center of the galaxy, dividing the galaxy
   centre (Fig.~\ref{Fig20}). }
\label{Fig3} 
\end{figure}

\subsection{Surface brightness profile}\label{Sec2.5}

We follow the reduction procedure by \citet[see references
therein]{2016MNRAS.462.3800D} and fit elliptical isophotes to the WFC3
$F110W$, ACS $F814W$ and SDSS $i$-band images of NGC~5322 using the
IRAF task {\sc ellipse} \citep{1987MNRAS.226..747J} to extract the
surface brightness, ellipticity ($\epsilon$ = 1 - b/a, where a and b
are the semi-major and minor-axes of the best-fit ellipse), position
angle (P.A.) and isophote shape parameter ($B_{4}$) profiles along the
major-axis (see Fig.~\ref{Fig2}b). $B_{4} $ quantifies deviations of
the isophotes from pure ellipses (i.e., $B_{4} =$ 0). If $B_{4} < $ 0
then the isophotes are boxy, $B_{4} >$ 0 if the isophotes are
discy. We run {\sc ellipse} on the WFC3 $F110W$, ACS $F814W$ and SDSS
$i$-band images using a mask for each image that is generated by
SExtractor \citep{1996A&AS..117..393B} together with a manual mask to
exclude the galaxy's nuclear dust disc, gaps between CCD detectors in
the images, background galaxies, bright foreground stars, and chip
defects.

Fig.~\ref{Fig3} shows a composite light profile of NGC~5322 together
with the pertaining $\epsilon$, P.A. and $B_{4}$ profiles, assembled
by combining our WFC3 $F110W$, ACS $F814W$ and SDSS $i$-band data. At
$R \la 3\arcsec$, we used the near-IR WFC3 $F110W$ light profile which
is significantly less contaminated by dust extinction (see
Sections~\ref{Sec3.2} and  \ref{Sec4.2.1}). In Fig.~\ref{Fig20}, we show the central
regions and associated masks of NGC~5322 to highlight the masked
nuclear dust regions in the WFC3 $F110W$ and ACS $F814W$ images.
The ACS $F814W$ data, with a better spatial resolution than the WFC3
and SDSS data, were used at $3\arcsec \la R \la 40\arcsec$. In order
to capture NGC~5322's stellar halo that extends beyond the FOVs of the
ACS and WFC3 images, at large radii ($R \ga 40\arcsec$) we used the
SDSS $i$-band data to constrain the sky-background. The WFC3 $F110W$
and SDSS $i$-band data were shifted to match the ACS $F814W$
data. Although the WFC3 $F110W$ filter covers a wavelength range
mostly redder than those of the ACS $F814W$- and SDSS $i$-band
filters, we find a good overlap between the ACS $F814W$, SDSS $i$ and
WFC3 $F110W$ light profiles over the $R \la 10\arcsec$ radial range,
when we exclude the dust-affected region. The magnitudes in this paper
are in the VEGA magnitude system.

 \section{Structural Analysis}\label{Sec3}

\subsection{Residual image and dust map}\label{Sec3.1}

Fig~\ref{Fig1}(b) shows the residual image of NGC~5322 generated by
subtracting the IRAF {\sc ellipse} model image from the ACS $F814W$
image. Four structural features are evidenced in this image: (i) a
dark nuclear disc along the galaxy's major axis which is due to the
edge-on inner dust disc (with a radius of 1$\farcs$7 $\approx$ 246.5
pc and a thickness of $0\farcs$6 $\approx$ 87 pc near the galaxy
centre, respectively) that is visible in the galaxy images crossing
the centre of the galaxy, (ii) two bright cones along the major axis,
(ii) two bright cones along the minor-axis and (iv) a large-scale
`X'-shaped structure due to the spheroid's boxy isophotes (see also
Fig.~\ref{Fig2}a and Section \ref{Sec3.3}). The dust disc, with
$F435W - F814W \sim$ 2.72$^{+0.13}_{-0.39}$ compared to the relatively
bluer immediate surroundings ($F435W - F814W \sim 2.30$), is better
visible in the colour maps (Figs.~\ref{Fig1}c and \ref{Fig2}b), rather
than in the residual image, as an inner disc with red colour excess of
$\sim 0.42$ mag (see also \citealt{1997ApJ...481..710C};
\citealt{2000AJ....120..123T}). To create this colour map, we
registered the $F435W$ and $F814W$ images using the IRAF tasks {\sc
  geomap} and {\sc gregister}. Excluding the dust disc, the galaxy
becomes progressively bluer toward larger radii, with an
$F435W - F814W$ colour gradient of $\sim 0.22$ mag per 10$\arcsec$. We
note that the dust disc has a thickness of $\sim 0\farcs$2 $\approx$
28 pc near the galaxy center in the WFC3 $F110W$ image
(Fig.~\ref{Fig20}).

\begin{table*}
\setlength{\tabcolsep}{0.06240in}

\begin {minipage}{180mm}
~~~~~~~~~~~\caption{Structural parameters for NGC 5322 resulting from
  the fit to the optical data }
\label{Tab2}
\begin{tabular}{@{}lllcccccccccccccccccccccccccccccccccccccccccccccc@{}}
\hline
\hline
Galaxy&Type &$ \mu_{\rm b} $ & $R _{\rm b}$ &$R_{\rm b}$ &$
                                                           \gamma$&$\alpha$&$n_{\rm
                                                                             cS}$&$R_{\rm
                                                                                   e,cS}$&$R_{\rm
                                                                                           e,cS}$&$\mu_{\rm
                                                                                                   e,S
                                                                                                   }
                                                                                                   $&$n_{S}$&$R_{\rm
                                                                                                              e,S}$&$R_{\rm
                                                                                                                     e,S}$&$\mu_{\rm
                                                                                                                            0,halo}$&$h$&$h$\\                                                                                                           
&& &(arcsec)&(pc)&&&&(arcsec)&(kpc)&&&(arcsec)&(pc)&&(arcsec)&(kpc)\\
(1)&(2)&(3)&(4)&(5)&(6)&(7)&(8)&(9)&(10)&(11)&(12)&(13)&(14)&(15)&(16)&(17)\\
\multicolumn{6}{c}{} \\ 
\hline
                              
 NGC~5322      &E3-4 & 14.36  &0.37  &53.7&0.17 &2   &4.0   &30.4  &4.4&17.21  &1.0&2.46&356.7&  20.79&53.6&7.8\\
&&&10\%&&10\%&&20\%&25\%&&&20\%&20\%&&&10\%&&\\
\hline
\hline
\end{tabular} 

Notes.--- First row: structural parameters from the three-component S\'ersic
inner disc + core-S\'ersic spheroid + exponential halo fit to the
$F814W$-band ($\sim I$-band) surface brightness profiles of NGC 5322
(Fig.~\ref{Fig3}). Col.\ 1: galaxy name. Col.\ 2: morphological
type. Cols.\ 3$-$10: best-fit parameters from the core-S\'ersic
model. Cols.\ $11- 14$: S\'ersic model parameters for the exponential
($n=1$) inner stellar disc with scale length
($h= R_{\rm e,S}/1.678) \sim 1\farcs47$ and central surface brightness
($\mu_{\rm 0,disc})\sim \mu_{\rm e,S} -1.822 \sim 15.39 $ mag
arcsec$^{-2}$. Cols.\ 15$-$17 best-fit parameters for the exponential
halo light.  $ \mu_{\rm b}$, $ \mu_{\rm e}$ and $\mu_{0,\rm halo}$ are
in mag arcsec$^{-2}$. The second row shows the uncertainties on the
fit parameters. The uncertainty  associated with $\mu_{\rm b}$,
$\mu_{\rm e,S}$ and $\mu_{\rm 0,h}$ is $\sim$ 0.02 mag arcsec$^{-2}$.
\end{minipage}
\end{table*}


\subsection{Structural decomposition} \label{Sec3.2}

NGC~5322 is well known for its inner (discy) stellar component, which
counter-rotates with respect to the galaxy's boxy spheroid (e.g.,
\citealt{1988A&A...202L...5B,1996IAUS..171..181B};
\citealt{1992A&A...258..250B}; \citealt{1992MNRAS.254..389R};
\citealt{1994A&AS..104..179G}; \citealt{1995A&A...293...20S};
\citealt{1997ApJ...481..710C}; \citealt{1998AJ....116.2793S}), and for
its central radio source (\citealt{1984A&A...137..362F};
\citealt{1984A&A...134..207H}). Analysis of high-resolution {\it HST}
ACS/WFC images shows that the galaxy contains a small-scale discy
stellar component, a boxy spheroid (that has a depleted core as
revealed by the flattened inner stellar distribution which deviates
downward relative to the inward extrapolation of the spheroid's outer
S\'ersic profile) and an outer stellar halo (Figs.~\ref{Fig1},
\ref{Fig2} and \ref{Fig3}).  As mentioned in the Introduction,
early-type galaxies brighter than $M_{B} \sim -20.5$ mag, hereafter
core-S\'ersic galaxies, have depleted cores and central stellar mass
deficits which are thought to be created through the ejection of stars
by inspiralling binary SMBHs (not due to central dust obscuration),
while those with $M_{B} \ga -20.5$ mag tend to be coreless (e.g.,
\citealt{1997AJ....114.1771F}; \citealt{1999ASPC..182..124K};
\citealt{2001AJ....121.2431R}; \citealt{2001AJ....122..653R};
\citealt{2004AJ....127.1917T}; \citealt{2005AJ....129.2138L,
  2007ApJ...662..808L}; \citealt{2006ApJS..164..334F};
\citealt{2012ApJ...755..163D, 2013ApJ...768...36D,
  2014MNRAS.444.2700D}; \citealt{2013AJ....146..160R}).

The light distributions of core-S\'ersic galaxies are well described
by the core-S\'ersic model\footnote{\citet{2012ApJ...755..163D,
    2014MNRAS.444.2700D} discussed the core-S\'ersic model in detail
  and provided a comparison between the model and the Nuker model
  \citep{1995AJ....110.2622L}, see also
  (\citealt{2003AJ....125.2951G}; \citealt{2004AJ....127.1917T};
  \citealt{2006ApJS..164..334F}).}  which is a blend of an inner
power-law and an outer S\'ersic model
\citep{2003AJ....125.2951G}. Detailed multicomponent decompositions
using high-resolution light profiles, extending to at least
$1R_{\rm e}$ (\citealt{2003AJ....125.2951G}), are crucial to
understand properly the different formation mechanisms that build up
the distinct structural components in galaxies.  Therefore, we fit a
three-component S\'ersic small-scale disc + core-S\'ersic spheroid +
exponential halo model to the major-axis surface brightness profile of
NGC~5322 which extends out 200$\arcsec$.  Fig.~\ref{Fig3} shows this
light profile decomposition with a small rms residual scatter of
$\sim 0.015$ mag arcsec$^{-2}$. Each model component was convolved
with a Gaussian PSF with FWHM $\sim 0\farcs25$ (see
\citealt{2017arXiv170702277D}, their Section~3.1). The FWHM of the PSF
was determined using several stars in the WFC3 F110W image of the
galaxy. Table~\ref{Tab2} provides the best-fitting model parameters.

We found that a S\'ersic model fit to the halo light yields
$n \sim 1.0 \pm 0.2$. We therefore force an exponential ($n=1$) model
for the outer stellar halo. Similarly, the embedded, small-scale discy
component was modelled using a S\'ersic model with a half-light
effective radius
$R_{\rm e} \sim 2\farcs46 \pm 0\farcs49 \approx 356.7 \pm 71$~pc and
$n \sim 1.0$, indicating an inner exponential disc with a scale length
$h \sim 1\farcs47 \pm 0\farcs15 \approx 212.6 \pm 21.3$~pc and an $I$-band
central surface brightness $\mu_{0,\rm disc} \sim 15.39 \pm 0.02$ mag
arcsec$^{-2}$. The boxy spheroid which dominates over
$R \la 100\arcsec$ has a core-S\'ersic stellar light distribution with
$n \sim 4.0 \pm 0.8$ (i.e., essentially the
\citet{1948AnAp...11..247D} $R^{1/4}$ law outside the depleted core),
$R_{\rm b} \sim 0\farcs37 \pm 0\farcs04$,
$\mu_{{\rm b},I} \sim 14.36 ~\pm ~0.02$ mag arcsec$^{-2}$ and
$R_{\rm e} \sim 30\farcs4 ~\pm ~ 7\farcs6$. At $R \ga 100\arcsec$, the
spheroid and outer stellar halo contribute equally.

\subsection{Literature comparison}

\begin{figure}
 \hspace*{-.10599cm}   
\vspace*{-.157299cm}   
 \includegraphics[angle=270,scale=0.425]{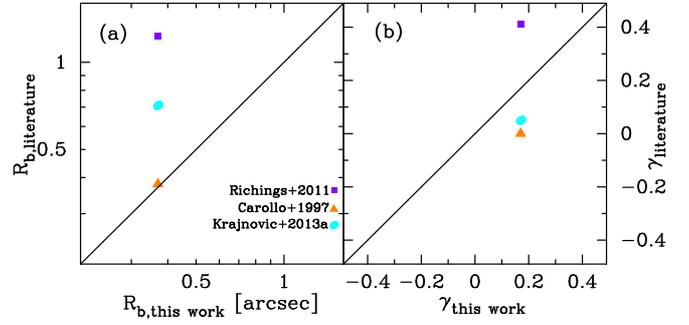}
 \caption{Comparison of our break radius ($R_{\rm b}$) and inner
   logarithmic slope of the spheroid's inner light profile ($-\gamma$)
   with those values from \citet{1997ApJ...481..710C},
   \citet{2011MNRAS.415.2158R} and \citet{2013KRb}. The solid lines
   show a one-to-one relation.}
\label{Fig5} 
\end{figure}

For comparison, \citet{1992MNRAS.254..389R} decomposed the radially
limited $\sim10\arcsec$ absorption line profile of NGC~5322, which
does not probe the outer parts of this galaxy's spheroid and stellar
halo, into an inner stellar disc with $h \sim 2\farcs1$ and a spheroid
with $R_{\rm e} \sim 25\arcsec$ using a double Gaussian model. Direct
comparison between our decomposition and that by
\citet{1992MNRAS.254..389R} is challenging given the different
approaches by these two works, i.e., absorption line profile vs. light
profile decomposition. \citet[their Figure A.9]{1995A&A...293...20S}
fit an exponential disc + $R^{1/4}$ bulge model to their
$\sim10\arcsec$ data with a seeing FWHM of 1$\farcs6$ and a plate scale of
0$\farcs46/$pixel, finding $R_{\rm e} \sim 3.4$ kpc for the bulge, and
$h \sim 180$ pc, a $V$-band $\mu_{0,\rm disc}\sim$ 19.45 mag
arcsec$^{-2}$.

Fig.~\ref{Fig5} compares our break radius ($R_{\rm b}$) and inner
logarithmic slope of the spheroid's inner light profile ($-\gamma$)
with those from \citet{1997ApJ...481..710C},
\citet{2011MNRAS.415.2158R} and
\citet{2013KRb}. \citet{1997ApJ...481..710C} and \citet{2013KRb} fit
the Nuker model (\citealt{1995AJ....110.2622L, 2005AJ....129.2138L})
to $10\arcsec$ {\it HST} WFPC2/PC1 light profile of NGC~5322, while
\citet{2011MNRAS.415.2158R} modelled the galaxy's {\it HST} ACS/WFC
light profile using the core-S\'ersic model. Also, \citet[their
Fig.~1]{2013KRa} modelled their ground based light profile of the
galaxy over the radius range 2\fc5 $\la R \la 172\arcsec$ using a
single S\'ersic model, resulting in a residual profile which clearly
reveals structures that have not been properly modelled. Because these
past works (\citealt{1997ApJ...481..710C};
\citealt{2011MNRAS.415.2158R}; \citealt{2013KRa,2013KRb}) treated the
galaxy as a single-component system, the fraction of the light
attributed to the spheroid in their modellings come from the inner
stellar disc and outer halo. This led, in part, the measurements of
the break radius/half-light radius and S\'ersic index by
\citet[$R_{\rm b}\sim$ 1\fc23, $n \sim 8.1$ and
$R_{\rm e} \sim 93\fc1$, their Table~2]{2011MNRAS.415.2158R} and
\citet[$n \sim 5.8$, $R_{\rm e} \sim 67\fc9$, their Fig.~1]{2013KRa}
to be biased high (Fig.~\ref{Fig5}).
Measuring the core size as the radius where the negative logarithmic
slope of the fitted Nuker model profile equals 1/2 ($R_{\gamma=1/2}$),
\citet[their Table C1]{2013KRb} found $R_{\gamma=1/2} \sim 0\fc71$,
while \citet[their Table 4]{1997ApJ...481..710C} reported
$R_{\gamma=1/2} \sim 0\fc38$ which agrees with our core-S\'ersic break
radius for the galaxy (Fig.~\ref{Fig5}a).

Part of the discrepancies between our work and
\citet{1997ApJ...481..710C}, \citet{2013KRb} and
\citet{2011MNRAS.415.2158R} arises from the treatment of the inner
dust disc. Indeed, the inner negative logarithmic slopes presented in
\citet[$\gamma = 0.0$]{1997ApJ...481..710C}, and
\citet[$\gamma = 0.05$]{2013KRb} are shallower than that of ours
($\gamma \sim 0.17 \pm 0.02$), while \citet{2011MNRAS.415.2158R}
reported a very steep inner negative logarithmic slope for the galaxy
($\gamma = 0.41$, Fig.~\ref{Fig5}b). Furthermore, replacing the inner
$R\la 3\arcsec$ profile in Fig.~\ref{Fig3}, extracted from the redder
WFC3 $F110W$ image, with the ACS $F814W$ data and performing a
3-component decomposition of the ACS $F814W$ ($R \la 40\arcsec$) +
SDSS $i$-band ($R > 40\arcsec$) profile yields a core with
$R_{\rm b} \sim 0\farcs49 \pm 0\farcs05 \approx 71.0 \pm 7.3$ pc,
compared to that adopted here
($R_{\rm b} \sim 0\farcs37 \pm 0\farcs04 \approx 53.7 \pm 5.4$ pc, see
Fig.~\ref{Fig3} and Table~\ref{Tab2}). In Section \ref{Sec4.2.1}, we
measure low $F110W$-band extinction ($A_{F110W} \sim 0.031 - 0.047$
mag) in the regions immediately surrounding the masked regions near
the galaxy centre (Figs.~\ref{Fig20} and \ref{Fig2}).
Also, we show that our break radius for this galaxy is in good
agreement with those predicted based on the
\citet{2014MNRAS.444.2700D} $R_{\rm b}-\sigma$ and $R_{\rm b}-M_{V}$
relations, i.e., $R_{\rm b, \sigma-based} \approx 44.0 \pm 6.1$ pc and
$R_{{\rm b},M_{V}-{\rm based}} \approx 48.6 \pm 7.2$ pc, respectively
(Section~\ref{Sec4.2}).  This $R_{\rm b}$ agreement, together with the
low IR extinction, gives us some confidence that our measurement of
NGC~5322's break radius was not affected by the nuclear dust disc.

\subsection{Ellipticity, Position Angle, Isophote Shape and Kinematics }\label{Sec3.3}

Inside $R \sim 0\farcs3$, NGC~5322 shows radial variations in
ellipticity, P.A. and $B_{4}$ due to the nuclear dust disc and the PSF
(Figs.~\ref{Fig1} and \ref{Fig3}). Going outward from
$R \sim 0\farcs3$, the ellipticity continuously rises from
$\epsilon \sim 0.10$ to a maximum of $\epsilon \sim 0.45$ at
$\sim 2\farcs5$ and it then declines reaching a local minimum of
$\sim 0.25$ at $\sim 10\arcsec$, before starting to gently rise
again. The ellipticity profile together with the positive $B_{4}$
values over $1\farcs5- 8\arcsec$ reveal the small-scale disc,
consistent with the 3-component profile decomposition
(Fig.~\ref{Fig3}). The spheroid of NGC~5322 contains boxy isophotes,
with ellipticity $\epsilon \sim 0.35 \pm 0.07$, that resulted in the
`X'-shaped structure in the residual image (Figs.~\ref{Fig1} and
\ref{Fig2}). 
\citet[see also \citealt{1992MNRAS.254..389R};
\citealt{1995A&A...293...20S} and
\citealt{2013KRa}]{1988A&A...202L...5B} found that the inner region
($R \la 2\farcs6$), where this galaxy's inner disc contributes
significantly, counter-rotates with maximum rotational velocity
$V_{\rm rot,max} \sim$ 80 km s$^{-1}$ relative to the intermediate
parts the spheroid at $10\arcsec \la R \la 35\arcsec$
($V_{\rm rot,max} \sim -$ 30 km s$^{-1}$). \citet[their
Fig.~9]{1992MNRAS.254..389R} reported
$V_{\rm rot,max}/\sigma \sim 1.4$ and $\sim 0.15$ for the inner disc
and spheroid, respectively.

The small-scale stellar disc, spheroid and outer stellar halo of
NGC~5322 are well aligned as revealed by the position angle 
which remains stable at $\sim 94^{\circ}$ outside the region most
affected by the PSF (see Fig.~\ref{Fig3}).

\subsection{Radio structure}\label{Sec3.4}

NGC~5322 is a radio source which hosts a LINER nucleus
\citep{2009A&A...508..603B}. Fig.~\ref{Fig1}(d) shows the 1.5 GHz
e-MERLIN radio continuum map of NGC~5322 with a restoring beam of
$\theta_{\rm maj} \times \theta_{\rm min} \sim 0\farcs43\times 0\farcs
17$ at a P.A. $\sim$ 0.79$^{\circ}$ (Table~\ref{Tab_radio}). Lower
resolution images of this source in the Faint Images of the Radio Sky
at Twenty Centimeters (FIRST) survey (\citealt{1995ApJ...450..559B})
at 1.4 GHz show that it has an extended radio jet structure along the
same position angle of the inner jet structure observed in Fig 1.
Whilst the VLA maps of the galaxy by \citet{1984A&A...134..207H} also
revealed that the galaxy has an FRI radio morphology with twin radio
jets extending over $\sim 20\arcsec$ at 1.4 GHz.  These results are
consistent but substantially shallower and at lower angular
resolution, as compared to these new e-MERLIN observations. While the
large bandwidth of these observations provides near complete aperture
coverage to emission on scales between 0.1 to several arcsec, it
should be noted that the shortest project baseline of e-MERLIN
($\sim 11$ km) in these observations means that these data are
insensitive to very extended, diffuse emission.  In the case of
NGC~5322 this spatial filtering acts to separate the central jet
emission from diffuse radio emission.  Fig.~\ref{Fig1}(d) shows that
these radio jets extend $\sim 10\farcs9 \approx 1.6$ kpc at a position
angle of 172$^{\circ}$ as observed with e-MERLIN. On large scales, the
radio jets show slight deviations with respect to the orientation of
the inner jet.

Fig.~\ref{Fig2} shows the radio contours overlaid on the {\it HST} ACS
residual image and the ACS $F435W - F 814W$ colour map. In
Fig.~\ref{Fig2}(b), we overplot the elliptical isophotes obtained from
the ellipse fit to the ACS F814W image to illustrate the position
angle of the galaxy relative to the nuclear dust disk and radio
jets. Because of NGC~5322's nuclear dust disc crossing the galaxy
centre, we did not attempt to align these images by shifting the radio
core to the central maximum in the ACS images. We determine that the
offset between the optical and radio images can be up to
$\sim 0\farcs3$. The inner jet is nearly perpendicular to the nuclear
dust and stellar discs that are aligned along the major axis of the
galaxy (i.e., P.A. $\sim$ 94$^{\circ}$). The radio jets do
not have strong enough optical/near-IR emissions to be seen from the
optical and near-IR images, or as excess optical/IR emission above the
(inner disc + spheroid + halo) model fit to the galaxy light profile
(Fig~\ref{Fig3}).
\begin{table*}
\setlength{\tabcolsep}{0.030356240in}

\begin {minipage}{180mm}
~~~~~~~~~~~\caption{\textrm{\normalfont{e}}-MERLIN observation}
\label{Tab_radio}
\begin{tabular}{@{}lllcccccccccccccccccccccccccccccccccccccccccccccc@{}}
\hline
\hline
Frequency &$\theta_{\rm maj} \times \theta_{\rm min}  $&Beam P.A.&$\theta_{\rm
                                                         maj} \times
                                                         \theta_{\rm
                                                         min} $&Beam P.A.
  &$S_{\rm peak}$ &$S_{\rm int,core}$ &$T_{\rm B,core}$&$P_{\rm core}$&$S_{\rm int,jets}$ &$P_{\rm jets}$&\\

(GHz)&&($^{\circ}$)&&($^{\circ}$)&(mJy beam$^{-1}$)&(mJy)&(K) &(W Hz$^{-1}$)&(mJy) &(W Hz$^{-1}$)&\\

(1)&(2)&(3)&(4)&(5)&(6)&(7)&(8)&(9)&(10)&(11)\\
\multicolumn{6}{c}{} \\ 
\hline
                              
 1.5   &$0\farcs43\times 0\farcs
17$& 0.79 &$0\farcs20\times 0\farcs04$ &$169.6 \pm 0.094 $&5.38$\pm
                                                              0.01$
                  &$5.96\pm0.02$  &4.5$\times 10^{7}$ &$6.55\times10^{20}$&$6.41\pm 0.02$ &$7.04\times10^{20}$ \\
\hline
\hline
\end{tabular}

Notes.--- Col. 1: centrally weighted (i.e., nominal) frequency. Col. 2: synthesised
beam major- and minor-axis dimensions. Col. 3: synthesised beam
position angle.  Cols. 4 and 5: deconvolved beam dimensions and
position angle obtained from a single Gaussian component fit to the
central radio emission using the {\sc aips} task {\sc jmfit}.  Col. 6:
{\sc jmfit} peak flux density of the core emission.  Col. 7: {\sc
  jmfit} integrated flux density of the core emission.  Cols. 8 and 9:
brightness temperature and power of the radio core. Col. 10:
integrated flux density of the jets. Col. 11: radio power of the
jets. The quoted errors on the P.A., $S_{\rm peak}$ and
$S_{\rm int,core}$ (Cols.~5, 6, and 7) are those reported by the {\sc
  jmfit}. The error on $S_{\rm int,jets}$ is that given by {\sc
  tvstat}.

\end{minipage}
\end{table*}

\subsubsection{Flux density, brightness temperature and radio power}

We fit a single elliptical Gaussian model to the central radio source
using the {\sc aips}\footnote{http://www.aips.nrao.edu/} task {\sc
  jmfit}, deriving the position, peak and integrated intensity flux
density of the core emission.  Also, the {\sc jmfit} fit yields
deconvolved beam dimensions and position angle after deconvolving the
clean beam from the fitted component size (see Table~\ref{Tab_radio}).

The measurement of the brightness temperature of a radio source is a
diagnostic tool that discriminates between the thermal free-free
emission from H {\sc ii} regions and non-thermal synchrotron radio
continuum emission. The former has a nebular brightness temperature of
$T_{\rm B} \sim 10^{4}$ K, while the latter has a higher brightness
temperature $T_{\rm B} \ga 10^{6} - \ 10^{11} $ K (e.g,
\citealt{1992ARA&A..30..575C}; \citealt{2001ApJS..133...77H}).

Using the Rayleigh-Jeans Law, the brightness temperature ($T_{\rm B}$)
can be written as

 \begin{equation}
T_{\rm B} =
 \left(\frac{S_{\nu}}{\Omega_{\rm beam}\nu^{2}}
 \right) \left(\frac{c^{2}}{2k}\right), 
\label{Eq1}
 \end{equation}

 where $\nu$ is the central frequency, $k$ is the Boltzmann constant,
 c is the speed of light, $\Omega_{\rm beam}$ is the deconvolved beam
 solid angle and $S_{\nu}$ is the integrated radio flux density. For
 the radio core of NGC 5322 we derived
 $T_{\rm B} \sim 4.5 \times 10^{7}$~K at $\nu$ $\sim$ 1524 MHz,
 suggesting a non-thermal radio continuum emission. This confirms the
 conclusion from VLA 15 GHz and VLBA observations presented by \citet
 {2005A&A...435..521N} who find a brightness temperature of
 $> 10^{8.2}$ K.

 We run the {\sc aips} task {\sc tvstat} to determine the integrated
 radio flux density for the jets
 $S_{\rm int,jets}\sim 6.41 \pm 0.02$~mJy, which yields a 1.5 GHz
 radio power of $\sim 7.04 \times~10^{20}$~W Hz$^{-1}$ at NGC~5322's
 distance of 30.3 Mpc (Table~\ref{Tab_radio}). Similarly, the core
 radio continuum emission of $S_{\rm int,core}\sim 5.96 \pm 0.02$~mJy
 for NGC~5322 implies a core radio power of
 $\sim 6.55 \times 10^{20}$~W Hz$^{-1}$. The NRAO VLA Sky Survey
 (NVSS, \citealt{1998AJ....115.1693C}), with a very poor resolution of
 45$\arcsec$, has detected a source, near NGC~5322's centre, with a
 1.4 GHz integrated flux of 78 $\pm$ 2.8 mJy. The FIRST survey finds
 three sources with integrated radio flux densities of $\sim$
 40.7~mJy, 8.8~mJy and 8.1~mJy at $\sim 3\farcs1$ (south), $7\farcs0$
 (north), and $11\farcs9$ (north) from the galaxy center,
 respectively. However, given the FIRST  images' resolution of
 $5\arcsec$, this survey's detection of the three sources and the
 pertaining flux values for NGC~5322 are not reliable.  Nonetheless,
 our total recovered flux density
 ($S_{\rm int,jets} + S_{\rm int,core} \sim 12.4$~mJy) is
 significantly lower ($\sim$factor of 5) than that recovered for more
 compact/filled interferometer observations (e.g., FIRST,
 $S_{\rm int,total}\sim 57.6$~mJy) due to primarily the limited number
 of short baselines in the e-MERLIN array which makes it insensitive
 to structures greater than $5\arcsec$. As such our estimate of the
 jet power for the galaxy is a lower limit.

\section{Discussion}\label{Sec4}

\subsection {Central mass deficit }\label{Sec4.1}

The central stellar mass deficits of core-S\'ersic galaxies are
thought to be created by inspiralling binary SMBHs that form from
`dry' galaxy mergers (see Section~\ref{Sec3.2}). Theory predicts that
the impact of multiple dry major mergers on core creation is
cumulative. N-body simulations by \citet{2006ApJ...648..976M}
suggested that the stellar mass deficit $M_{\rm def}$ that is
generated by binary SMBHs after $\mathcal{N}$ successive mergers
equals $0.5 \mathcal{N}  M_{\rm BH}$, where $M_{\rm BH}$ is the total sum of the
masses of the binary SMBHs. A careful measurement of the central
stellar mass deficit in the spheroid of NGC~5322 can be used to
constrain the galaxy's merger history.  We follow the procedures by
\citet[their Section~5.2]{2014MNRAS.444.2700D} and compute the stellar
luminosity deficit ($L_{\rm def}$) as the difference in luminosity
between the inwardly-extrapolated outer S\'ersic profile of the
complete core-S\'ersic model (fit to the spheroid's profile) and the
core-S\'ersic model (see Section~\ref{Sec3.2} and
Fig.~\ref{Fig3}). This yields stellar luminosity deficit in the
$I$-band\footnote{We use an $I$-band absolute magnitude for the Sun of
  $M_{\sun,I} \sim 4.14$ mag
  (http://www.ucolick.org/~cnaw/sun.html).},
$L_{\rm def}/ L_{\sun,I} \sim 3.9 \times 10^{8}$. To convert this
luminosity deficit into $M_{\rm def}$, we determined the stellar
mass-to-luminosity ratio ($M/L)$ excluding the inner dust disc and
using the central dust-free colour of the galaxy
($F435W - F814W \sim2.15$, Fig~\ref{Fig1}), which corresponds to
$V-I \sim 1.24$ \citep {1995PASP..107..945F}.  For $V-I \sim 1.24$,
the colour-age-metallicity-($M/L$) diagram by \citet[their
Fig.~A1]{2009MNRAS.397.2148G} yields an $M/L \sim 4.2$ in the
$V$-band, which corresponds to an $I$-band $M/L\sim$ 2.8 \citep[his
Table 5A]{1994ApJS...95..107W}, implying
$M_{\rm def} \sim (1.1 \pm 0.3)\times 10^{9} M_{\sun}$.

\subsection {Merger rate}\label{Sec4.2}

Dynamical SMBH measurements are not available for NGC~5322.  We
therefore compute the $M_{\rm def}/M_{\rm BH} $ ratios using SMBH
masses predicted from the $M_{\rm BH}-\sigma$ and $M_{\rm BH}-L$
relations. For the velocity dispersion of NGC~5322 ($\sigma$ $\sim$
229 km~s$^{-1}$), the \citet{2013ApJ...764..151G} $M_{\rm BH}-\sigma$
relation predicts log~$(M_{\rm BH}/M_{\sun}) \sim 8.51 \pm 0.40$. To
estimate the total $I$-band absolute magnitude of the the spheroid
($M_{{\rm sph}, I}$), we integrate the core-S\'ersic profile
(Fig.~\ref{Fig3}) by taking into account the spheroid's average
ellipticity ($\epsilon \sim 0.35$). This yields
$M_{{\rm sph}, I} \sim -23.04$ mag, corrected for Galactic extinction
($\sim -$ 0.02 mag) and surface brightness dimming ($\sim -$ 0.03
mag), which corresponds to a $V$-band magnitude
$M_{{\rm sph}, V} \sim -21.77$ mag applying the colour transformation
$V-I \sim 1.27$ \citep{1995PASP..107..945F}.  Converting the
\citet{2013ApJ...764..151G} $B$-band core-S\'ersic $M_{\rm BH}-L$
relation into a $V$-band relation using $B-V$ = 0.95
\citep{1995PASP..107..945F}, we predict 
log~$(M_{\rm BH}/M_{\sun}) \sim 8.93 \pm 0.34$ for the spheroid of
NGC~5322.

We thus find that $M_{\rm def}/M_{\rm BH} \sim$ 3.4 and 1.3 when using
SMBH masses predicted based on the spheroid's $\sigma$ and $L$,
respectively. While these numbers suggest that the spheroid has
experienced $\mathcal{N} \sim 2 - 7$ major mergers
\citep{2006ApJ...648..976M}, this conclusion is hinged on the
assumption that SMBH binary scouring is the sole mechanism that
created the core. For comparison, past works have found mass deficits
that are typically $\sim 0.5 - 4~M_{\rm BH}$
(\citealt{2004ApJ...613L..33G}; \citealt{2004AJ....127.1917T};
\citealt{2006ApJS..164..334F}; \citealt{2013AJ....146..160R};
\citealt{2013ApJ...768...36D, 2014MNRAS.444.2700D}). If the central
mass deficit of NGC~5322 has been enlarged by additional mechanisms
(such as the scouring action of gravitationally recoiled SMBH that
oscillates about the centre of a merger remnant, e.g.,
\citealt{2004ApJ...613L..37B}; \citealt{2008ApJ...678..780G}, and the
action of multiple SMBHs that form in galaxies at high redshifts,
e.g., \citealt{2012MNRAS.422.1306K}, see the discussion by
\citealt[their Sections 5.3 and 5.4]{2014MNRAS.444.2700D}), then
$\mathcal{N} \la 2 - 7$.

\subsubsection{Dust absorption}\label{Sec4.2.1}

Nuclear dust absorption can reduce the surface brightnesses of
NGC~5322 at the inner regions, creating an artificial depleted core or
enlarging the break radius/central mass deficit and resulting in
$\mathcal{N} \la 2 - 7$ for the galaxy. However, not only did we
carefully mask the inner dust disc to extract the WFC3 IR $F110W$
profile at $R\la 3\arcsec$ (see Sections \ref{Sec2.5}, \ref{Sec3.1}
and Fig.~\ref{Fig20}) but also the values of $\sigma \sim 229$ km
s$^{-1}$ and $M_{V} \sim -21.77$ mag for NGC~5322's spheroid imply
that the \citet[their Table~3]{2014MNRAS.444.2700D} $R_{\rm b}-\sigma$
and $R_{\rm b}-M_{V}$ relations predict break radii of
$R_{\rm b, \sigma-based} \approx 44.0$ pc and
$R_{{\rm b},M_{V}-{\rm based}} \approx 48.6$ pc, in good agreement
with the break radius from our modelling (Fig.~\ref{Fig3},
Table~\ref{Tab2}).

Below, as a further check on the effect of the dust absorption, we
follow the dust correction steps described in \citet[their Appendix A,
see also \citealt{2015ApJ...807...56B}, their
Section~3.3]{2008MNRAS.391.1629N} and measure the $F110W$
($\sim J$-band) extinction ($A_{J}$) in the nuclear regions of the
galaxy using our $F435W - F814W$ ($B-I$) colour map (Figs.~\ref{Fig1}
and \ref{Fig2}).

Assuming that dust acts as a foreground screen obscuring
the stars, the extinction in the $I$-band can be written as
\begin{equation}
A_{I}= I_{obs} - I_{corr} = \alpha_{\rm red} E(B-I),
\label{Eq2}
 \end{equation}
 where $I_{obs}$ and $I_{corr}$ are the observed and
 extinction-corrected $I$-band surface magnitudes.  The reddening
 $E(B-I)= (B-I)_{obs} - (B-I)_{corr}$ and
 $\alpha_{\rm red} = ((A_{B}/A_{I})-1)^{-1} $.  Therefore,
 \begin{equation}
f_{I,corr}= f_{I,obs} \bigg(\frac{f_{I,obs}}{f_{B,obs}}\bigg)^{\alpha_{\rm red}}
\bigg(\frac{f_{B,corr}}{f_{I,corr}}\bigg) ^{\alpha_{\rm red}}, 
\label{Eq3}
 \end{equation}
where $f_{I}$ and $f_{B}$ are the $I$- and $B$-band surface fluxes.

Assuming a very shallow stellar population gradient in the central
region of NGC~5322 (\citealt{2010MNRAS.408..272S}), we found the dust
free surface flux ratio in the inner regions to be
$f_{B,corr}/f_{I,corr} \sim 0.125$. Following \citet[their Table
6]{2011ApJ...737..103S}, $A_{\rm ACS,F475W} = 3.610 E(B - V)$,
$A_{\rm ACS,F814W} = 1.526 E(B - V)$ and
$A_{\rm WFC3,F110W} = 0.881 E(B - V)$ for $R_{V}=3.1$. Thus,
$A_{\rm ACS,F475W}/A_{\rm ACS,814W} \sim 2.366$, i.e.,
$\alpha_{\rm red} \sim 0.732$ and
$A_{\rm WFC3,F110W}/A_{\rm ACS,814W} \sim 0.5773$.  Therefore,
combining eqs. \ref{Eq2} and \ref{Eq3} gives an $I$-band extinction
value of $A_{I} \sim 0.080$ mag for the regions, near the galaxy
centre, immediately outside the ACS and WFC3 masked regions
(Fig.~{\ref{Fig20}}) with a $B-I$ colour of $\sim$2.37 mag. This
$A_{I} $ value corresponds to a low IR extinction value of
$A_{J} \sim 0.047$ mag.  For comparison, $A_{I} \sim 0.053$ mag for
the outer part of the dust disc (as traced by the colour map) near the
galaxy centre with a $B-I$ colour of $\sim$2.33 mag. For the reddest
part of the dust disc with $B-I$ $\sim$ 2.85 mag, which we have
masked, $A_{I} \sim 0.053$ mag (see Figs.~\ref{Fig1}, \ref{Fig20} and
\ref{Fig2}). Thus, we find a low IR extinction value of
$A_{J} \sim 0.031$ mag for the outer part of the dust disc, while for
the reddest part $A_{J} \sim 0.250$ mag. These three $A_{J}$ values
can be compared to our adopted uncertainty on surface brightnesses
$\mu_{\rm b}, \mu_{\rm e}$ and $\mu_{\rm 0}$, i.e., $\sim$ 0.02 mag
arcsec$^{-2}$.

We did not attempt to derive a dust-corrected $F110W$ surface
brightness profile for NGC~5322 by creating a dust-corrected WFC3
$F110W$ image of the galaxy using eq. \ref{Eq3}. Although the
foreground screen approximation (eq. \ref{Eq2}) enables us to
constrain the uncertainties due to the effects of the nuclear dust,
the technique is not fully reliable to derive a robust light profile
for the galaxy (\citealt{2008MNRAS.391.1629N};
\citealt{2015ApJ...807...56B}). Both \citet[Appendix
A]{2008MNRAS.391.1629N} and \citet[Section 3.3]{2015ApJ...807...56B}
have estimated the best $\alpha_{\rm red}$ values in a subjective
manner by iterating $\alpha_{\rm red}$ to obtain what they considered
the best dust-corrected images for their galaxies.

\subsection{Spheroid, inner disc and outer halo formation and the
  connection to the central radio source}\label{Sec4.4}

Here we discuss the formation of NGC~5322, an E$3-4$ giant elliptical
in a poor group (\citealt{1993A&AS..100...47G};
\citealt{2004ApJ...607..810M}), considering the implications of our
findings; a galaxy spheroid with a stellar mass deficit
$M_{\rm def} \sim (1.1 \pm 0.3)\times 10^{9} M_{\sun}$, small-scale
stellar and dust discs, an outer stellar halo and a central radio
source (Section~\ref{Sec3.4}, see also \citealt{1984A&A...137..362F};
\citealt{1984A&A...134..207H}).

\subsubsection{Spheroid property}\label{Sec4.4.1}

In the previous sections, we have discussed that the depleted core in
NGC~5322 is not due to the nuclear dust disc crossing the galaxy
center, which can create an artificial core or enlarge a pre-existing
depleted core. However, it is still worth comparing the properties of
the spheroid of NGC~5322 with those of core-S\'ersic galaxies to
ensure that our core identification is
real. \citet{1997AJ....114.1771F} noted the connection between the
``core'' vs. ``power-law'' central structural dichotomy and the
``boxy'' vs. ``discy'' and ``slow rotator'' vs. ``fast rotator''
divides (see also \citealt{1983ApJ...266...41D};
\citealt{1987ApJ...312..514C}; \citealt{1988A&A...202L...5B};
\citealt{1988A&AS...74..385B}; \citealt{1990AJ....100.1091P};
\citealt{1994AJ....108.1567J}). In agreement with this
picture\footnote{A caveat here is that the ``core''
  (\citealt{1995AJ....110.2622L}; \citealt{1997AJ....114.1771F}) and
  ``core-S\'ersic'' (\citealt{2003AJ....125.2951G}) structural
  classifications are not identical.  Unlike ``core-S\'ersic''
  galaxies, ``core'' galaxies identified by the Nuker model (e.g.,
  \citealt{1995AJ....110.2622L}) do not necessarily contain
  partially-depleted cores relative to their spheroids' outer S\'ersic
  profile \citep{2012ApJ...755..163D, 2013ApJ...768...36D,
    2014MNRAS.444.2700D, 2015ApJ...798...55D}.}, NGC~5322's
core-S\'ersic spheroid rotates slowly (e.g.,
\citealt{1988A&A...202L...5B}; \citealt{1992MNRAS.254..389R};
\citealt{2011MNRAS.414..888E}; \citealt{2013KRb,2013KRa}) and has boxy
isophotes (Figs.\ 2 and 3). The spheroid's S\'ersic ($n \sim 4$)
stellar light distribution outside the core, $M_{V} \sim -21.77$ mag
$\approx 73.4\%$ of the total galaxy light (i.e., including the
stellar halo) and $\sigma \sim$ 229 km s$^{-1}$, also suggest that the
galaxy is a core-S\'ersic galaxy. Moreover, there is a tendency for
the AGN of core-S\'ersic galaxies to be more radio-loud than those of
the coreless galaxies (e.g., \citealt{2005A&A...440...73C,
  2006A&A...453...27C}; \citealt{2009ApJS..182..216K};
\citealt{2011MNRAS.415.2158R}). Indeed, NGC~5322 is classified as a
radio-loud galaxy (e.g., \citealt{2009A&A...508..603B}).

\subsubsection{Kinematically decoupled core}

The gas-poor major merger scenario of the spheroid
(Section~\ref{Sec4.4.1}) cannot naturally account for the galaxy's
inner stellar disc which counter-rotates rapidly with respect to the
slowly rotating spheroid (\citealt{1988A&A...202L...5B}, see Section
\ref{Sec3.3}). Galaxies with central regions that are kinematically
decoupled from the rest of the galaxy are thought to be common, but
their formation mechanisms are still unclear (e.g.,
\citealt{1982MNRAS.201..975E}; \citealt{1988A&A...202L...5B};
\citealt{1988ApJ...327L..55F}; \citealt{1991Natur.354..210H};
\citealt{1990ApJ...361..381B}; \citealt{1995AJ....109.1988F};
\citealt{1997ApJ...481..710C}; \citealt{2006MNRAS.373..906M};
\citealt{2015ApJ...802L...3T}). Integrating our S\'ersic fit and using
$\epsilon \sim 0.4$, we find an $I$-band $M_{I}\sim-$ 19.29 mag for
the counterrotating disc, dominated at all radii by the spheroid
(Fig.~\ref{Fig3}). This disc only makes up $\approx 2.6\%$ of the
total galaxy light and using a mass-to-light ratio $M/L_{I} \sim 2.8$,
the same as the spheroid's $M/L_{I}$, gives a stellar disc mass of
$M_{*,\rm disc} \sim$ (6.4 $\pm$ 1.8)$\times$ 10$^{9}$ $M_{\sun}$,
somewhat smaller than the stellar disc mass from
\citet[$M_{*,\rm disc} \sim
10^{10}M_{\sun}$]{1990dig..book..232B}. The residual structure along
the major axis of the galaxy due to the counterrotating stellar disc
(Figs.~\ref{Fig1} and \ref{Fig2}) is in excellent agreement with the
galaxy velocity map by \citet[their Fig.~C4]{2011MNRAS.414.2923K}.

Kinematically decoupled core (KDC) formation due to survived cores of
accreted gas-poor satellites has been discussed by
\citet{1984ApJ...287..577K} and \citet{1990ApJ...361..381B}. These
mechanisms are, however, inconsistent with this galaxy's inner dust
disc and central radio source which are a telltale signature of
nuclear cold gas accretion events. It is unlikely that the KDC of
NGC~5322 (i.e., the inner stellar disc) was built through major
mergers of gas-rich or partially-gaseous galaxies (e.g.,
\citealt{1991ApJ...370L..65B}; \citealt{1991Natur.354..210H};
\citealt{1992A&A...258..250B}; \citealt{2006MNRAS.373..906M};
\citealt{2011MNRAS.416.1654B}; \citealt{2011MNRAS.414.2923K};
\citealt{2014MNRAS.444.3357N}; \citealt{2014ApJ...783..120T};
\citealt{2015MNRAS.452....2K}; \citealt{2015ApJ...802L...3T}).  These
processes drive nuclear inflows of gas and subsequently trigger
intense central starbursts, which are typically $\sim 1\%-5\%$ of the
host galaxy light (\citealt{2008ApJ...679..156H, 2009ApJS..181..486H},
see also \citealt{2017ApJ...845..128P}), resulting in a coreless
galaxy (e.g., \citealt{1997AJ....114.1771F}). Furthermore, using
numerical simulations, \citet{1994MNRAS.270L..23H} proposed a
formation model for KDC galaxies, including NGC~5322, where a
retrograde flyby interaction of a galaxy with another elliptical
galaxy leads to a slowly rotating halo and a rapidly rotating KDC
core. Their simulations assume an inner core plus an other halo
components for the galaxies, but NGC~5322 has an inner stellar disc, a
spheroid, and an outer halo (See Fig.~\ref{Fig3} and
Section~\ref{Sec3}).

The properties of NGC~5322  favour the formation of the KDC due to
the accretion of a gas-rich satellite which settles at the galaxy centre.

\subsubsection{A three-phase
assembly for NGC~5322}

We argue that a more likely scenario for NGC~5322 is a three-phase
assembly, in which, the galaxy had its spheroid built earlier through
($\mathcal{N} \sim 2-7$) violent dry major mergers involving SMBHs
(e.g., \citealt{1997AJ....114.1771F}; \citealt{1999ASPC..182..124K};
\citealt{2001AJ....121.2431R}; \citealt{2001AJ....122..653R};
\citealt{2004AJ....127.1917T}; \citealt{2005AJ....129.2138L,
  2007ApJ...662..808L}; \citealt{2006ApJ...640..241B};
\citealt{2006ApJS..164..334F}; \citealt{2012ApJ...747...85X};
\citealt{2012ApJ...744...85M, 2016ApJ...830...89M}), followed by the
subsequent accretion of (a satellite with) metal-rich gaseous material
which settles at the galaxy centre into a rapidly rotating (discy)
cold gas reservoir. This, in turn, later fuels the central radio
source and converts into the inner counter-rotating stellar disc (see
\citealt{1988Natur.335..705B}; \citealt{1988ApJ...327L..55F};
\citealt{1997AJ....114.1771F}; \citealt{2000ApJS..127...39C};
\citealt{2002ApJ...578..787C}). \citet{1992MNRAS.254..389R} argued
that this stellar disc is embedded in the dynamically hot spheroid
(i.e., $V_{\rm rot}/\sigma \sim 1.4$).  The outer halo
($\approx$24.0\% of the total galaxy light) could naturally have grown
later via dry minor mergers and the accumulation of stars stripped
during flybys (e.g., \citealt{2009ApJ...702.1058Z};
\citealt{2012MNRAS.425.3119H, 2013MNRAS.429.2924H};
\citealt{2015MNRAS.449..528H}; \citealt{2016arXiv160909498R}, their
Fig.~4). Due to NGC~5322's group environment and therefore the lack of
intracluster light (\citealt{1993A&AS..100...47G};
\citealt{2004ApJ...607..810M}), this galaxy's halo and spheroid
contribute equally in the outermost regions, compared to BCG halos
which typically dominate their host spheroids' light at large radii
(e.g., \citealt{2007MNRAS.378.1575S}; \citealt{2011ApJS..195...15D};
\citealt{2017arXiv170702277D}).

\subsubsection{Low-luminosity AGN feedback}

A key challenge to our three-phase formation picture is that the new
star formation due to the gas/gas-rich satellite accretion can refill
the depleted core (see \citealt{1997AJ....114.1771F}).  However, the
core can be protected if the non-thermal AGN feedback swiftly quenches
the in-situ star formation after the stellar disc is formed.
\citet{2012MNRAS.422.1835S} detected neutral hydrogen absorption in
NGC~5322, although \citet{2003AJ....125..667H} did not detect neutral
hydrogen associated with the galaxy. Using e-MERLIN observations, we
tentatively detected neutral hydrogen absorption near the central
region of NGC 5322. Also, \citet{2010MNRAS.408..272S} found a
significant $\alpha$ enhancement at the central regions of the galaxy
([$\alpha/\rm Fe$] $\sim0.30$), suggesting short star-formation
timescales (e.g., \citealt{1987A&A...185...51M};
\citealt{1999MNRAS.302..537T, 2005ApJ...621..673T}).

Our radio data analysis suggests that NGC~5322 houses a low-luminosity
AGN (LLAGN).  As noted in Section~\ref{Sec3.4}, the 1.5 GHz radio
continuum emission of the galaxy has a core radio power and brightness
temperature of $P_{\rm core} \sim 6.55 \times 10^{20}$~W Hz$^{-1}$ and
$T_{\rm B,core} \sim 4.5 \times 10^{7}$~K, respectively
(Table~\ref{Tab_radio}), which agree with those those reported for
low-luminosity AGN ($T_{\rm B,LLAGN} \ga 10^{7}$~K e.g.,
\citealt{2002A&A...392...53N}; \citealt{2005A&A...435..521N};
\citealt{2013ApJ...779..173N} and
$P_{\rm core,LLAGN}\sim 10^{18}- 10^{25}$~W Hz$^{-1}$, e.g.,
\citealt{2001ApJS..133...77H}). For NGC~5322's radio jets, we measured
a low radio power of
$P_{\rm jets}\sim 7.04 \times 10^{20}$~W~Hz$^{-1}$.  Using Chandra
X-ray spectra and the surface brightness profile of NGC~5322,
\citet{2008A&A...486..119B} argued that a low accretion rate
($\approx$ 0.75 $M{\sun}$ yr$^{-1}$) of hot gas onto the AGN at the
bondi radius of $15$ pc is enough to account for these radio jets
kinetic power.  Fig.~\ref{Fig2} reveals that the residual image
obtained by subtracting the {\sc ellipse} model image of the galaxy
exhibits two bright cones, tracing the orientations of the radio
jets. However, since the jets have a low radio power, these bright
cones are less likely to form due to radio mode AGN photoionisation of
the interstellar medium of the galaxy (e.g.,
\citealt{2006MNRAS.365...11C}; \citealt{2006MNRAS.373L..65H};
\citealt{2009ApJ...698..594M}).

Furthermore, the modelling of the {\it HST} brightness profile shows
that NGC~5322 has a normal size core for the galaxy's $\sigma$ and
$M_{V}$, thus it is unlikely that a pre-existing large core in this
galaxy was partially replenished by new stars (see
Section~\ref{Sec4.2}).


\subsubsection{Three-phase scenario versus galaxy properties and
  environment}

The three-phase formation scenario reconciles well with the colour,
stellar population properties, globular cluster system and the
environment of NGC~5322. \citet[their Tables~C1, C2 and
C3]{2010MNRAS.408..272S} reported a positive age gradient for the
galaxy where the age rises from a central value of 7.4~Gyr to an
average age of 12.6~Gyr. The central intermediate age (7.4~Gyr)
stellar population (see also \citealt{2002MNRAS.333..517P}) is
consistent with a stellar population that is a mixture of relatively
young and old stars associated mostly to the stellar disc (i.e., KDC)
and spheroid, respectively. NGC~5322's negative radial metallicity
gradient of $\sim-$0.20 dex decade$^{-1}$ is such that the galaxy is
metal-rich in the central regions ([Z/H] \sm0.45] and it becomes less
metal-rich with increasing radius \citep{2010MNRAS.408..272S}.  This
negative metallicity gradient is expected since the small (gas-poor)
satellites forming the outer halo are metal-poor. The higher central
metallicity may reflect an early dissipative merger-induced starburst
in the (coreless) progenitors of the spheroid and may typically
remain intact during dry mergers (e.g.,
\citealt{2006MNRAS.366..499D}; \citealt{2010MNRAS.408..272S}) and the
accretion of (a satellite with) already enriched gaseous material that
built up the stellar disc. The galaxy becomes gradually bluer towards
larger radii (see Section~\ref{Sec3.1}), as also reflected in the
decline of the fraction of red globular clusters going from the centre
to larger radii (\citealt{2006ApJ...636...90H}, their Table~5 and
Fig.~24). In line with the three-phase picture, NGC~5322 resides in a
poor galaxy group environment, which favours mergers (e.g.,
\citealt{1998ApJ...496...39Z}; \citealt{2001ApJ...563..736C}).

The nuclear dust disc and the inner stellar disc of NGC~5322 are both
oriented along the major-axis of the galaxy, suggesting a common
(external) origin for both these structures. The dust disc is nearly
perpendicular to the radio jets (Section~\ref{Sec3.4}). However,
\citet{2005A&A...435...43V} did not find a trend between the radio jets
and dust orientations for their sample of galaxies with nuclear dust
disc, in contrast they found that radio jets tend to be perpendicular
to the dust structure in the `dust lane' galaxies.  A number of past
studies found a strong link between the presence of dust in galaxies
and radio activities (e.g., \citealt{1994AJ....108.1567J};
\citealt{1995AJ....110.2027V}; \citealt{1999AJ....118.2592V};
\citealt{1999ApJS..122...81M}; \citealt{2001AJ....121.2928T};
\citealt{2005A&A...435...43V}; \citealt{2005AJ....129.2138L};
\citealt{2008A&A...489..989B}; \citealt{2012MNRAS.423...59S}). The
dust disc in NGC~5322 may signal the accretion of externally acquired
gaseous material onto the the central supermassive black hole which
feeds the AGN and powers the radio jets, extending $\sim 1.6$ kpc at a
position angle of 172\dg.

\section{Conclusions}\label{Conc}

As part of the LeMMINGs survey, we investigated the nuclear activity
and central structure of the radio galaxy NGC~5322 using a new
sub-arcsec L-band (1.25 - 1.75 GHz) e-MERLIN radio observation
together with {\it HST} WFC3/IR, ACS/WFC and SDSS imaging. We
extracted a composite ({\it HST} WFC3/IR and ACS/WFC + SDSS)
major-axis surface brightness profile and isophotal parameters.  The
inner $(R\la3\arcsec)$ part of this composite light profile was
extracted from the redder {\it HST} WFC3/IR image, applying a pixel
mask which mapped the galaxy's nuclear dust disc with a radius of
$1\farcs7$ $\approx$ 246.5 pc and a thickness of $0\farcs$6 $\approx$
87 pc near the galaxy center in the $F814W$/ACS/WFC image. For the
first time we have decomposed NGC~5322's composite light profile,
covering $R \sim 200\arcsec$, into an inner S\'ersic
($n\sim 1 \pm 0.2$) stellar disc, a boxy core-S\'ersic spheroid plus
an outer, exponential stellar halo. Our primary findings are follows:

(1) We have derived a central stellar mass deficit (\mdef~\sm 1.1
$\times 10^{9}$\sn) in the core-S\'ersic spheroid with a core size
$R_{\rm b}$ \sm 0\fc37 $\pm $ $0\farcs04$ $\approx$ 53.7 $\pm$ 5.4 pc
and $n \sim 4 $ $\pm$ 0.8. This core size is in good agreement with
those estimated utilising the spheroid's $\sigma \sim 229$ km s$^{-1}$
and $M_{V} \sim -21.77$ mag ($\approx 73.4\%$ of the total galaxy
light) along with the $R_{\rm b}-\sigma$ and $R_{\rm b}-M_{V}$
relations for core-S\'ersic galaxies \citep{2014MNRAS.444.2700D},
i.e., $R_{\rm b, \sigma-based} \approx 44.0 \pm 6.1$ pc and
$R_{{\rm b},M_{V}-{\rm based}} \approx 48.6 \pm 7.2$ pc. We also
measured low $F110W$-band extinction ($A_{F110W} \sim 0.031 - 0.047$
mag) in the regions immediately surrounding the masked regions near
the galaxy centre, compared to our adopted uncertainty on the surface
brightnesses $\mu_{\rm b}$, $\mu_{\rm e,S}$ and $\mu_{\rm 0,h}$, i.e.,
$\sim$ 0.02 mag arcsec$^{-2}$.  The good agreement in $R_{\rm b}$ and
the low IR extinction give us confidence that our measurement of
NGC~5322's $R_{\rm b}$ (and therefore $M_{\rm def}$) is not plagued by
the effects of the nuclear dust disc. We found a stellar mass \mstar
~\sm $(6.4 \pm 1.8)\times 10^{9}$\sn~ for the stellar disc, somewhat
smaller than that reported in the past.

(2) We find that NGC~5322 has a core 1.5 GHz radio continuum emission
with a power of $P_{\rm core} \sim 6.55 \times 10^{20}$~W Hz$^{-1}$
and a brightness temperature of
$T_{\rm B,core} \sim 4.5 \times 10^{7}$~K
(Table~\ref{Tab_radio}). Both these figures favour a non-thermal
nuclear radio emission due to a low-luminosity AGN. For the galaxy's
radio jets, we measured a low radio power of
$P_{\rm jets}\sim 7.04 \times 10^{20}$~W~Hz$^{-1}$.

(3) We propose a three-phase formation scenario to explain the
formation and evolution of NGC~5322. The spheroid may have built up
early through ($2-7$) gas-poor major mergers, as traced by the
$M_{\rm def}/M_{\rm BH} $ ratio. The galaxy subsequently 
cannibalised a satellite containing metal-rich gas, with stars and gas
settling at the centre and forming the rapidly rotating stellar disc,
counterrotating with respect to the spheroid. The outer stellar halo
later grew inside-out via minor mergers and accretion of
tidal debris. Our formation model for NGC~5322 is consistent with the
galaxy's colour, stellar population properties, globular cluster system
and environment.

(4) NGC~5322's dust disc oriented along the galaxy's major-axis is
nearly perpendicular to the radio jets, which extend $\sim 1.6$ kpc at
a position angle of 172\dg. Nuclear dust features may indicate the
accretion of externally acquired gaseous material onto the AGN,
powering central radio sources. For NGC~5322, this radio mode
(low-luminosity) AGN feedback that drives the low power jets might
have quenched the late-time star formation in the galaxy after the
stellar disc is formed, preventing the refill of the depleted core.

\section{ACKNOWLEDGMENTS}
We thank the referee for their careful reading of the paper and many
constructive suggestions that improved the paper.  BTD acknowledges
support from a Spanish postdoctoral fellowship ``Ayudas para la
atracci\'on del talento investigador. Modalidad 2: j\'ovenes
investigadores, financiadas por la Comunidad de Madrid'' under grant
number 2016-T2/TIC-2039. BTD acknowledges financial support from the
Spanish Ministry of Economy and Competitiveness (MINECO) under grant
number AYA2016-75808-R.  BTD \& JHK acknowledge financial support from
MINECO under grant number AYA2013-41243-P.  JHK acknowledges financial
support from the European Union’s Horizon 2020 research and innovation
programme under Marie Skłodowska-Curie grant agreement No 721463 to
the SUNDIAL ITN network, and from the Spanish Ministry of Economy and
Competitiveness (MINECO) under grant number AYA2016-76219-P. IMcH
thanks the Royal Society for the award of a Royal Society Leverhulme
Trust Senior Research Fellowship. RDB and IMcH also acknowledge the
support of STFC under grant [ST/M001326/1]. Based on observations made
with the NASA/ESA Hubble Space Telescope, and obtained from the Hubble
Legacy Archive, which is a collaboration between the Space Telescope
Science Institute (STScI/NASA), the Space Telescope European
Coordinating Facility (ST-ECF/ESA) and the Canadian Astronomy Data
Centre (CADC/NRC/CSA). e-MERLIN is a National Facility operated by the
University of Manchester at Jodrell Bank Observatory on behalf of
STFC. Funding for the Sloan Digital Sky Survey IV has been provided by
the Alfred P. Sloan Foundation, the U.S. Department of Energy Office
of Science, and the Participating Institutions. SDSS-IV acknowledges
support and resources from the Center for High-Performance Computing
at the University of Utah. The SDSS web site is www.sdss.org.

SDSS-IV is managed by the Astrophysical Research Consortium for the
Participating Institutions of the SDSS Collaboration including the
Brazilian Participation Group, the Carnegie Institution for Science,
Carnegie Mellon University, the Chilean Participation Group, the
French Participation Group, Harvard-Smithsonian Center for
Astrophysics, Instituto de Astrof\'isica de Canarias, The Johns
Hopkins University, Kavli Institute for the Physics and Mathematics of
the Universe (IPMU) / University of Tokyo, Lawrence Berkeley National
Laboratory, Leibniz Institut f\"ur Astrophysik Potsdam (AIP),
Max-Planck-Institut f\"ur Astronomie (MPIA Heidelberg),
Max-Planck-Institut f\"ur Astrophysik (MPA Garching),
Max-Planck-Institut f\"ur Extraterrestrische Physik (MPE), National
Astronomical Observatories of China, New Mexico State University, New
York University, University of Notre Dame, Observat\'ario Nacional /
MCTI, The Ohio State University, Pennsylvania State University,
Shanghai Astronomical Observatory, United Kingdom Participation Group,
Universidad Nacional Aut\'onoma de M\'exico, University of Arizona,
University of Colorado Boulder, University of Oxford, University of
Portsmouth, University of Utah, University of Virginia, University of
Washington, University of Wisconsin, Vanderbilt University, and Yale
University.

\bibliographystyle{mnras}

\label{lastpage}
\end{document}